\documentclass[aps,prmaterials,superscriptaddress,preprint,onecolumn,floatfix]{revtex4-2}

\usepackage[T1]{fontenc}
\usepackage[utf8]{inputenc}
\usepackage{multirow}
\usepackage{graphicx}
\usepackage{amsfonts,amsmath,amssymb,amsthm}
\usepackage{epstopdf}
\usepackage{upgreek,xspace}
\usepackage{chngcntr}
\usepackage{xr}
\usepackage{ulem}
\usepackage[colorlinks,allcolors=blue]{hyperref}
\usepackage[version=3]{mhchem}
\usepackage{booktabs}
\usepackage{longtable}
\usepackage{array}
\usepackage{tabularx}
\usepackage{newunicodechar}

\newunicodechar{₂}{$_2$}
\providecommand{\doi}[1]{\href{https://doi.org/#1}{\nolinkurl{#1}}}

\begin{document}
\bibliographystyle{apsrev4-2}

\title{Solid-State Dealloying Enables Local Symmetry Breaking in Ternary Intermetallic Thin Films}

\author{Mizuki~Ohno}
\affiliation{Department of Applied Physics and Materials Science, California Institute of Technology, Pasadena, California 91125, United States}
\affiliation{Institute for Quantum Information and Matter, California Institute of Technology, Pasadena, California 91125, United States}

\author{Reiley~Dorrian}
\affiliation{Department of Applied Physics and Materials Science, California Institute of Technology, Pasadena, California 91125, United States}
\affiliation{Institute for Quantum Information and Matter, California Institute of Technology, Pasadena, California 91125, United States}

\author{Veronica~Show}
\affiliation{Department of Applied Physics and Materials Science, California Institute of Technology, Pasadena, California 91125, United States}
\affiliation{Institute for Quantum Information and Matter, California Institute of Technology, Pasadena, California 91125, United States}

\author{Salva~Salmani-Rezaie}
\affiliation{Department of Materials Science and Engineering, The Ohio State University, Columbus, Ohio 43210, United States}

\author{Joseph~Falson}
\email{falson@caltech.edu}
\affiliation{Department of Applied Physics and Materials Science, California Institute of Technology, Pasadena, California 91125, United States}
\affiliation{Institute for Quantum Information and Matter, California Institute of Technology, Pasadena, California 91125, United States}
\begin{abstract}
Metastable quantum materials often occupy narrow composition windows with local symmetries distinct from competing equilibrium structures. These features open pathways for realizing qualitatively new electronic properties within a similar chemical subspace while at the same time complicating their deterministic synthesis. Here we illustrate a post-growth solid-state dealloying process using epitaxial ternary thin films in the La--Ag--Ge chemical space to realize a diffraction-averaged centrosymmetric superconducting structure. The process converts polar $P6_3mc$-LaAgGe into diffraction-averaged centrosymmetric AlB$_2$-type $P6/mmm$ La--Ag--Ge through net Ag loss during annealing at 800--900~$^\circ$C. Atomic-resolution electron microscopy reveals local Ag--Ge displacements of both signs relative to the planar configuration, consistent with local inversion-symmetry-breaking distortions. The converted films exhibit composition-dependent superconductivity with critical temperatures below 1~K and in-plane upper critical fields that exceed the weak-coupling Pauli-field estimate by a factor of approximately 6 in the thinnest superconducting films. These results establish epitaxial solid-state dealloying as a route to phase-selective synthesis of metastable phases, offering control over atomic-scale structural configurations and their interplay with emergent electronic properties.
\end{abstract}

\maketitle

Realizing quantum materials as epitaxial thin films requires simultaneous control of phase selection, stoichiometry, and nanoscale structural order.\cite{Samarth2017_QuantumMaterials} While epitaxial relationships, governed by film--substrate lattice commensurability and growth kinetics, can stabilize polymorphs outside their bulk equilibrium range,\cite{Gorbenko2002_Epitaxial, Zhou2024_Epitaxial} post-growth solid-state transformations provide a complementary strategy by separating precursor formation from subsequent chemical equilibration. These approaches include annealing a deposited precursor to crystallize it,\cite{Fujita2015_Oddparityb, Uchida2017_Quantum, Ohno2023_Novela, Ohno2023_Impactb} annealing in the presence of an auxiliary chemical reservoir to control the chemical potential of a volatile constituent,\cite{Fu2002_Superconducting, Kang2001_MgB2, Zhang2024_Peculiar} and topochemical reactions that selectively remove, insert, or exchange a mobile constituent while retaining part of the precursor framework.\cite{Gopalakrishnan1995_ChimieDouce, Li2019_Superconductivity, GutierrezLlorente2024_Topochemical, Zhang2014_Reversible, Zhou2016_CobaltChalcogenides, Averyanov2023_Class} These precedents suggest that sequential synthesis is especially well suited to multicomponent films combining a structurally persistent framework, one selectively mobile constituent, and a target phase with a narrow or metastable stability window.
Here we use solid-state dealloying to connect the structural and electronic properties of ternary intermetallic La--Ag--Ge thin films, with a focus on the sensitivity of superconductivity to local symmetry and dimensional confinement.\cite{Frigeri2004_Superconductivity, Maruyama2012_Locally, Llanos2026_Fieldinduced, Dorrian2026_PairBreaking}

The AlB$_2$-type structure provides a compelling test case in this ternary system. It consists of alternating triangular metal layers and planar honeycomb $X_2$ layers, and it hosts superconductors ranging from MgB$_2$, with a superconducting critical temperature $T_{\mathrm{SC}}=39$~K, to ternary CaAlSi and CaGaSi.\cite{Nagamatsu2001_Superconductivity, Imai2001_Superconductivity, Evans2009_Structural, Kortus2001_Superconductivity, An2001_Superconductivity} Replacing the honeycomb site with a mixed transition-metal/Ge sublattice generates a broad family of rare-earth compounds, $RE_2TX_3$ ($RE$ = rare earth, $T$ = transition metal, $X$ = Ge or Si), several of which superconduct below 4~K.\cite{Chen2012_Superconductivity, Kito2002_Superconductivity, Majumdar2001_Observation, Swiatek2024_Detailed, Ghosh2003_Superconductivity, Freccero2023_Flux, Hor2006_Superconductivity} Yet their phase stability is narrow: in La(Au$_x$Ge$_{1-x}$)$_2$, for example, the AlB$_2$ structure occurs only for $x\approx0.27$--0.375 within a sequence of competing ThSi$_2$, AlB$_2$, LiGaGe, and KHg$_2$ structure types.\cite{Peterson2023_Twists} These materials have consequently been studied mainly as polycrystalline bulk samples, where mixed-site disorder and off-stoichiometry are common and anisotropic transport or local structural correlations are difficult to access directly.\cite{Swiatek2024_Detailed}
Epitaxial studies of the related polar hexagonal $ABC$ compounds LaAuGe and LaPtSb have nevertheless demonstrated molecular-beam epitaxy (MBE) on Al$_2$O$_3$(0001), scanning transmission electron microscopy (STEM)-resolved layer buckling, and metallic transport.\cite{Du2019_Higha} Within the same structural family, the buckling is sensitive to electron count, chemical pressure, and lanthanide substitution, as shown in epitaxial LaAuSb and Gd$_x$La$_{1-x}$PtSb films.\cite{Strohbeen2019_Electronicallya, Du2024_Tunable}

Here we use a polar precursor as an epitaxial template and then transform it by solid-state dealloying. By accessing temperature ranges where Ag atoms are selectively desorbed in ultrahigh vacuum, polar LiGaGe-type LaAgGe ($P6_3mc$), grown on Al$_2$O$_3$(0001), converts into a diffraction-averaged centrosymmetric AlB$_2$-type La--Ag--Ge phase ($P6/mmm$). The converted phase occupies a bounded composition window, yet its lattice parameters, carrier response, and superconducting transition can be tuned by composition and thickness. Atomic-resolution imaging reveals local Ag--Ge displacements of both signs relative to the planar configuration, consistent with local inversion-symmetry-breaking distortions. The films superconduct below 1~K and show in-plane upper critical fields well above the weak-coupling Pauli-field estimate. By combining phase-selective synthesis, atomic-resolution imaging, and anisotropic transport, this work establishes a thin-film platform for examining how nanoscale local symmetry and dimensional confinement coexist with superconductivity.

\section{Results}

\subsection{Solid-state control of AlB$_2$-type phase stabilization}
\label{sec:dealloying}
\begin{figure}
  \centering
  \includegraphics[width=0.95\linewidth]{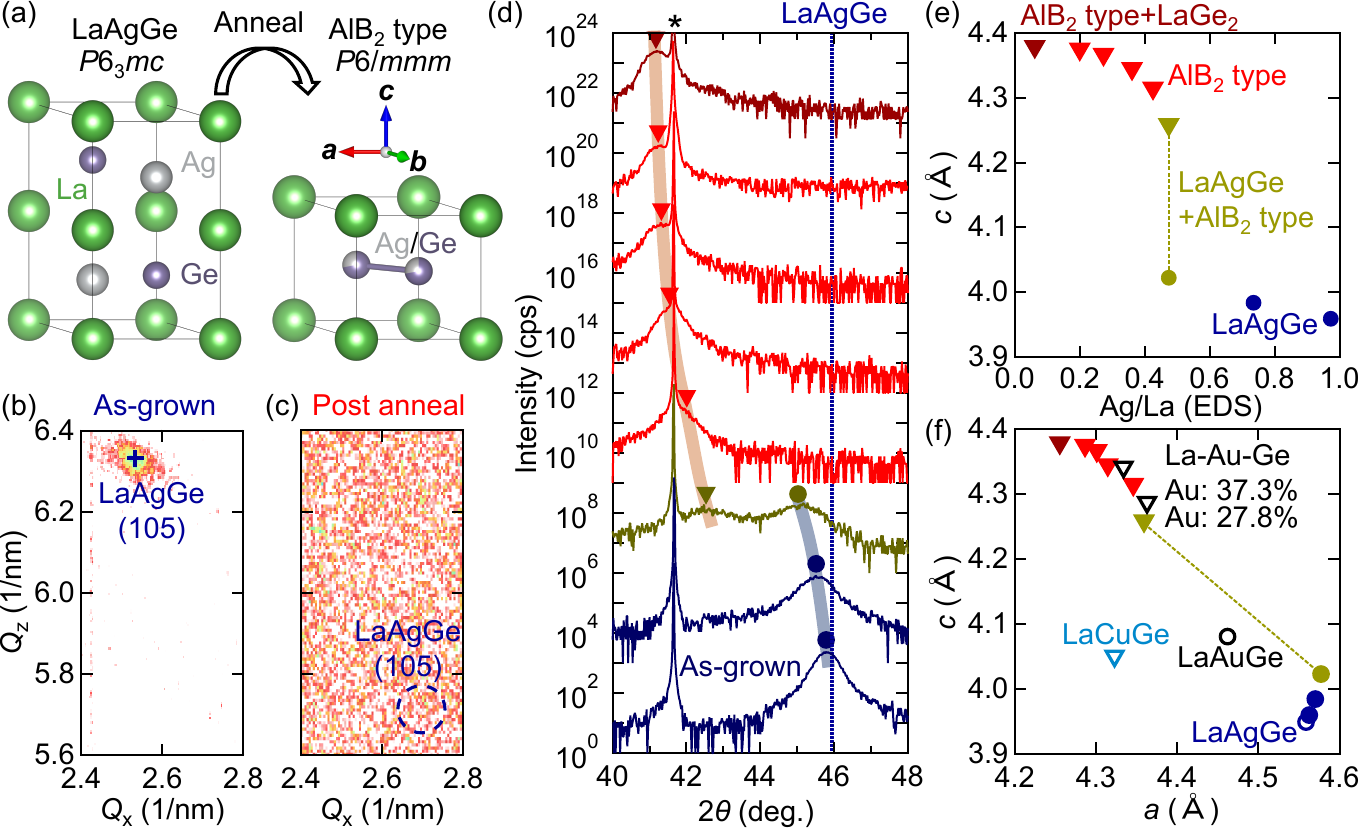}
  \caption{\textbf{Epitaxial solid-state dealloying toward AlB$_2$-type La--Ag--Ge.}
  (a) Crystal structures of polar LaAgGe ($P6_3mc$) and diffraction-averaged centrosymmetric AlB$_2$-type La--Ag--Ge ($P6/mmm$). Annealing drives net Ag loss and converts the ordered, polar La/Ag/Ge stacking into a diffraction-averaged planar honeycomb layer.
  (b,c) Reciprocal-space maps around the (105) reflection for (b) as-grown LaAgGe, showing a sharp diffraction spot, and (c) annealed AlB$_2$-type film, showing no measurable intensity at the same position, consistent with the transition from doubled-$c$ $P6_3mc$ to single-layer $P6/mmm$.
  (d) $\theta$--$2\theta$ X-ray diffraction (XRD) scans of films annealed at 800--900~$^\circ$C and an unannealed film, spanning Ag-poor AlB$_2$+LaGe$_2$ (top) to Ag-rich as-grown LaAgGe (bottom).
  Arrows track the peak shift. Al$_2$O$_3$ substrate peaks are marked with an asterisk.
  (e) Out-of-plane lattice parameter $c$ ($c/2$ for $P6_3mc$) versus the Ag/La ratio measured by energy-dispersive X-ray spectroscopy (EDS), with the relevant phases identified.
  (f) In-plane and out-of-plane lattice parameters ($a$, $c$; $c/2$ for $P6_3mc$) from the LaAgGe (106)/AlB$_2$-type (103) film peaks compared with bulk La(Au$_x$Ge$_{1-x}$)$_2$,\cite{Peterson2023_Twists} LaCuGe,\cite{Rieger1969_Ternaere} LaAuGe,\cite{Schnelle1997_Crystal} and LaAgGe.\cite{Pecharskii1991_ChemInform}
  Circles and triangles denote polar and diffraction-averaged centrosymmetric structures, respectively.
  }
  \label{fig:1}
\end{figure}
Figure~\ref{fig:1}a summarizes the sequential synthesis strategy. We use MBE to first establish an epitaxial LaAgGe template on Al$_2$O$_3$(0001) at a substrate temperature of $T_\mathrm{g}=600$~$^\circ$C. The as-grown film adopts the polar LiGaGe-type structure ($P6_3mc$), in which Ag and Ge occupy distinct Wyckoff sites and buckle along the $c$ axis. Across the explored flux ratios $P_\mathrm{Ag}/P_\mathrm{Ge}=0.6$--1.2, decreasing $P_\mathrm{Ag}/P_\mathrm{Ge}$ produces increasingly Ag-deficient LaAgGe and, at the most Ge-rich condition, an additional LaGe$_x$ phase. Even the as-grown LaAgGe films with longer out-of-plane repeats than the LaAgGe points in the annealed composition series remain polar LaAgGe and show no AlB$_2$-type reflection (Fig.~S1). Ag deficiency alone therefore does not stabilize the AlB$_2$-type ($P6/mmm$) phase during deposition; post-growth annealing is indispensable. After annealing, Ag and Ge nominally share a planar honeycomb site. Annealing is performed at 800--900~$^\circ$C under continual Ag supply, which limits excessive Ag desorption but still produces net Ag loss relative to the LaAgGe parent composition, as quantified by \textit{ex situ} EDS (Fig.~S5). Reciprocal space maps (RSMs) show a well-defined (105) superstructure reflection associated with the doubled-$c$ $P6_3mc$ stacking in the as-grown LaAgGe film (Fig.~\ref{fig:1}b) and no measurable intensity at the same position after conversion (Fig.~\ref{fig:1}c), consistent with collapse of the doubled-$c$ unit cell into the single-layer $P6/mmm$ AlB$_2$-type cell.

Qualitative and quantitative changes in the structure are also reflected in the out-of-plane $2\theta$ diffraction scans, as shown in Fig.~\ref{fig:1}d. The films were prepared using different combinations of precursor BEP ratio $P_\mathrm{Ag}/P_\mathrm{La}$, annealing temperature, and annealing time within the 800--900~$^\circ$C anneal window; they therefore do not constitute a monotonic temperature or time series. These parameters jointly determine the final EDS-measured Ag/La ratio and the corresponding phase and lattice parameters observed by XRD. The out-of-plane repeat obtained from these data ($c/2$ for $P6_3mc$ LaAgGe and $c$ for the AlB$_2$-type phase) is presented in Fig.~\ref{fig:1}e as a function of the Ag/La composition measured by EDS. The peaks at the highest diffraction angle in Fig.~\ref{fig:1}d, corresponding to the smallest out-of-plane repeat in Fig.~\ref{fig:1}e, originate from the as-grown, unannealed LaAgGe film. EDS gives Ag/La~=~0.97, establishing a nearly 1:1 Ag/La ratio, whereas XRD independently identifies the $P6_3mc$ LaAgGe phase but does not by itself determine stoichiometry. Relative to bulk LaAgGe, the as-grown film shows small increases of 0.09\% in $a$ and 0.27\% in $c/2$ (Fig.~\ref{fig:1}f).\cite{Pecharskii1991_ChemInform} This shift is in the same direction as the further lattice expansion observed for the Ag-poorer LaAgGe film at Ag/La~=~0.74 and is therefore consistent with a slight Ag deficiency in the as-grown film. Because epitaxial strain and other thin-film effects can also perturb the lattice parameters, however, the composition is assigned from EDS rather than from the lattice constants alone. The (00$l$) family of reflections shifts systematically with composition, tracking a monotonic expansion of the plotted out-of-plane repeat from 3.959~\AA\ (Ag/La = 0.97) to 4.379~\AA\ (Ag/La = 0.06) (Fig.~\ref{fig:1}e). Four distinct composition regions emerge across this parameter space: the parent LaAgGe phase for Ag/La $\gtrsim$ 0.7, a LaAgGe + AlB$_2$-type two-phase region containing the Ag/La~=~0.47 sample, a single-phase AlB$_2$-type region for Ag/La~=~0.20--0.43, and an AlB$_2$-type + LaGe$_2$ two-phase region containing the Ag/La~=~0.06 sample. RSMs around the Al$_2$O$_3$ $(1010)$ substrate reflection and the LaAgGe (106)/AlB$_2$-type (103) film reflections, together with azimuthal $\varphi$-scans confirming six-fold in-plane symmetry and a single, well-defined epitaxial registry with the substrate, are provided across the composition series in Fig.~S3.

The in-plane and out-of-plane lattice parameters place the AlB$_2$-type La--Ag--Ge phase on the same $a$--$c$ trend line as the reported bulk La(Au$_x$Ge$_{1-x}$)$_2$ series and the related LaAuGe and LaCuGe compounds (Fig.~\ref{fig:1}f).\cite{Peterson2023_Twists, Schnelle1997_Crystal, Rieger1969_Ternaere, Pecharskii1991_ChemInform}
Notably, the transformation from polar LaAgGe to diffraction-averaged centrosymmetric AlB$_2$-type La--Ag--Ge is accompanied by an in-plane lattice contraction and an out-of-plane expansion. This evolution is consistent with the geometric correlation between planar and puckered honeycomb layers established for bulk AlB$_2$-type superconductors.\cite{Evans2009_Structural}

\subsection{Atomic-scale structure of the annealed AlB$_2$-type films}
\label{sec:domains}

\begin{figure}
  \centering
  \includegraphics[width=0.95\linewidth]{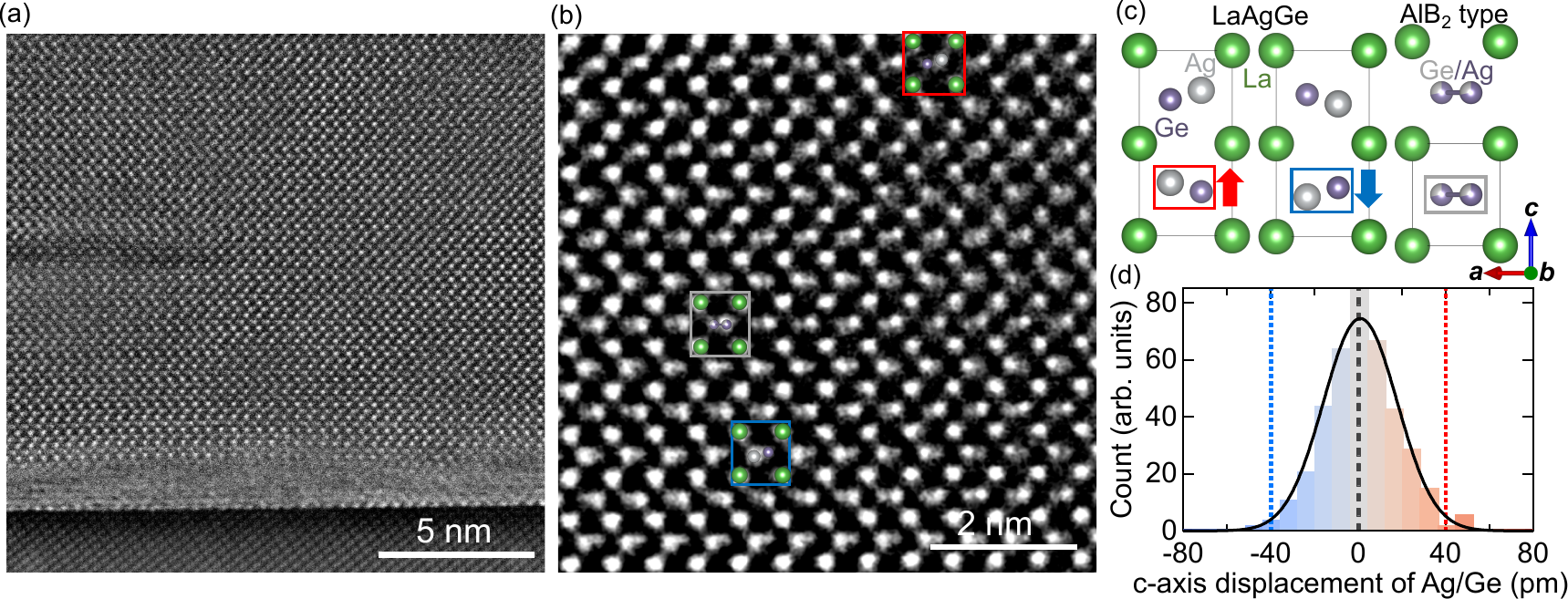}
  \caption{\textbf{Atomic-resolution STEM resolves local Ag--Ge displacements consistent with inversion-symmetry-breaking structural distortions in the AlB$_2$-type film.}
  (a) Wide-field cross-sectional high-angle annular dark-field scanning transmission electron microscopy (HAADF-STEM) image of the heterostructure.
  (b) Higher-magnification HAADF-STEM image resolving La and honeycomb-site Ag/Ge columns. Schematic structural models are overlaid on representative regions: the red and blue boxes indicate LaAgGe-like configurations with opposite signs of the Ag--Ge displacement, whereas the gray box indicates a nearly planar AlB$_2$-type configuration.
  (c) Reference structural models defining positive (red, upward arrow) and negative (blue, downward arrow) Ag--Ge $c$-axis displacements in polar LaAgGe and the zero-displacement planar AlB$_2$-type reference.
  (d) Histogram of the signed Ag--Ge $c$-axis displacement. The bars are colored from blue (negative) through gray (near zero) to red (positive), and the black curve represents a single-Gaussian fit. The gray dashed line marks zero displacement, and the blue and red dotted lines mark the LaAgGe reference displacements of $\pm39$~pm. Data are from one representative film with Ag/La~$\approx$~0.20.
  }
  \label{fig:2}
\end{figure}

The diffraction-averaged $P6/mmm$ assignment does not by itself establish whether the lattice is locally planar. Cross-sectional HAADF-STEM of an annealed film with Ag/La~$\approx$~0.20 confirms continuous film coverage and resolves the film/substrate interface (Fig.~\ref{fig:2}a). The higher-magnification image resolves the La and honeycomb-site columns (Fig.~\ref{fig:2}b). Three structural models are overlaid on representative regions to illustrate the observed local configurations: the red and blue overlays show LaAgGe-like arrangements with oppositely directed Ag--Ge displacements, whereas the gray overlay shows a nearly planar AlB$_2$-type arrangement. Because the HAADF intensity increases with atomic number, the brighter honeycomb-site columns are assigned as Ag-rich and the dimmer columns as Ge-rich. The signed $c$-axis displacement is defined from the relative positions of adjacent Ag-rich and Ge-rich columns, with the positive, negative, and zero-displacement reference configurations summarized in Fig.~\ref{fig:2}c.

The measured signed Ag--Ge displacements span both displacement directions relative to the planar AlB$_2$-type reference (Fig.~\ref{fig:2}d). The observed deviations from the planar configuration are consistent with local inversion-symmetry-breaking structural distortions. The measured displacements indicate spatial variation in the magnitude and direction of the local distortion, ranging from nearly planar to more strongly displaced configurations. Local FFT analysis reveals regions that retain the parent-like $c$-axis unit-cell doubling and regions in which the doubling is absent (Fig.~S6), providing additional evidence for spatially heterogeneous local structure. The displacement analysis was obtained from the atomic-resolution region shown here and is intended to characterize the range of local configurations observed within this field of view rather than to establish a statistically representative distribution across the film. Because the microscopy was performed on one representative composition (Ag/La~$\approx$~0.20), the composition dependence of these local structural variations is not determined here.

\subsection{Normal-state transport analysis}
\label{sec:normalstate}
\begin{figure}
  \centering
  \includegraphics[width=0.95\linewidth]{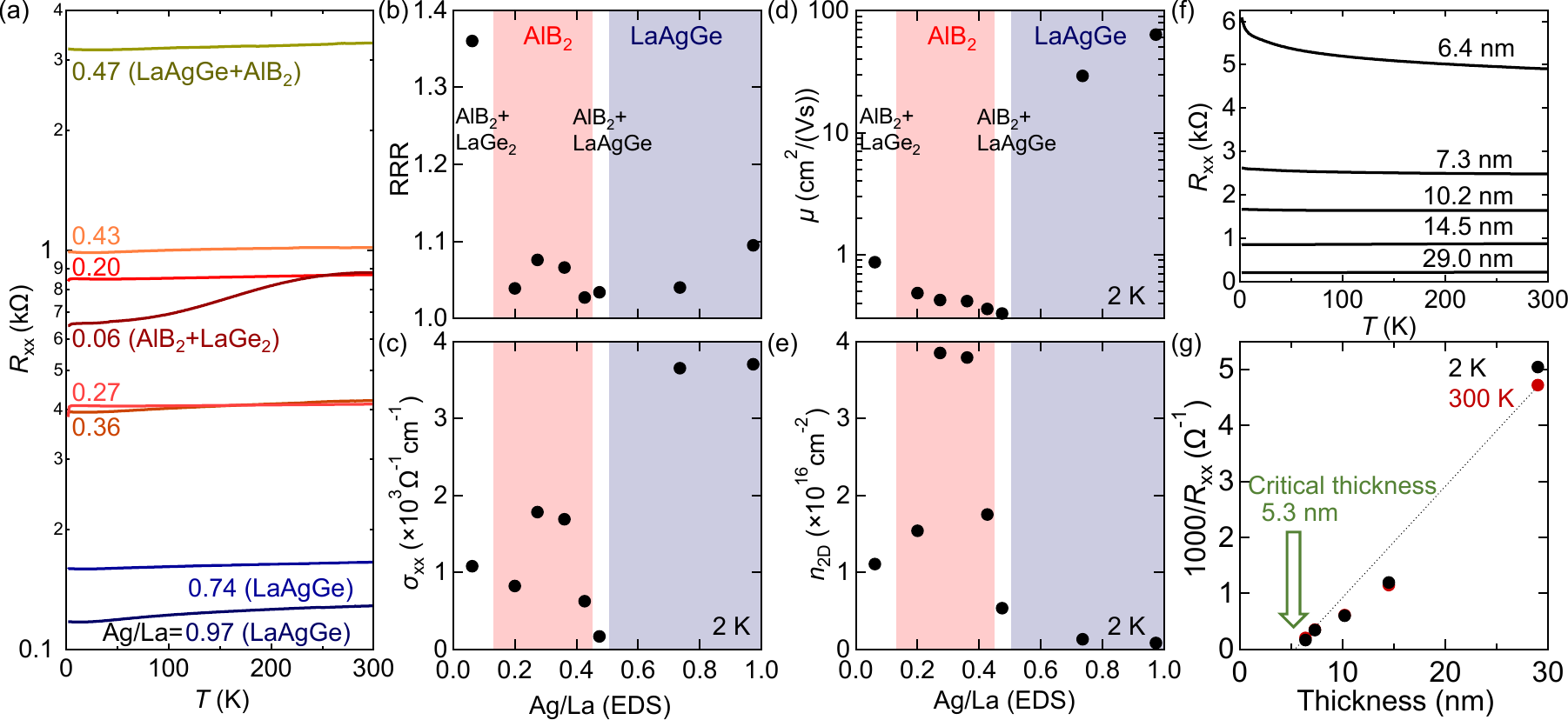}
  \caption{\textbf{Normal-state transport across the composition and thickness series.}
  (a) Temperature-dependent sheet resistance $R_\mathrm{xx}(T)$ for the composition series, spanning the single-phase AlB$_2$-type window (Ag/La = 0.20--0.43), the adjacent phase-boundary samples (Ag/La = 0.06 and 0.47), and the end-member LaAgGe phase (Ag/La = 0.74, 0.97).
  Ag/La (EDS) dependence of (b) the residual resistivity ratio (RRR), (c) the longitudinal conductivity $\sigma_\mathrm{xx}$, (d) the Hall mobility $\mu$, and (e) the effective sheet carrier density $n_\mathrm{2D}$ (single-band estimate) at 2~K. Shaded bands mark the AlB$_2$-type and LaAgGe single-phase regions, separated by the AlB$_2$+LaGe$_2$ and AlB$_2$+LaAgGe two-phase windows. RRR and $\sigma_\mathrm{xx}$ are both reduced toward the edges of the single-phase window and largest near its center.
  (f) Temperature-dependent sheet resistance $R_\mathrm{xx}(T)$ for the film-thickness series at fixed, Ag-deficient composition (Ag/La~$\approx$~0.20), for thicknesses from 6.4 to 29.0~nm.
  (g) Inverse sheet resistance $1000/R_\mathrm{xx}$ at 2~K and 300~K versus film thickness. The linear extrapolation to $1/R_\mathrm{xx}=0$ yields a critical thickness of $\approx$5.3~nm.
  }
  \label{fig:3}
\end{figure}

Figure~\ref{fig:3}b--e shows the composition dependence of the residual resistivity ratio (RRR), longitudinal conductivity $\sigma_\mathrm{xx}$, Hall mobility $\mu$, and effective sheet carrier density $n_\mathrm{2D}$ obtained from a single-band analysis of $R_\mathrm{yx}(B)$ at $T=2$~K. Because the electronic structure may be multiband and composition dependent, $n_\mathrm{2D}$ is best interpreted as an effective transport parameter rather than a direct measure of a unique Fermi-surface carrier density. Reporting the sheet density avoids introducing a film-thickness normalization into the carrier-density data. RRR remains close to unity throughout the single-phase AlB$_2$-type window, indicating that temperature-independent elastic scattering dominates over phonon scattering at all measured compositions. Drude estimates give mean free paths $\ell\approx0.16$--0.26~nm across the single-phase composition series (Table~S1), smaller than the lattice constant and consistent with a strongly disordered transport regime. Across the films in the thickness series at fixed Ag/La~$\approx$~0.20, the estimated mean free path spans $\ell\approx0.04$--0.4~nm (Table~S2).

The full field dependence of the magnetoresistance ratio and Hall resistance is provided for the composition and thickness series in Figs.~S7 and~S8. LaAgGe and the single-phase AlB$_2$-type films show a hole-like Hall response, whereas the Ag/La~=~0.06 film containing LaGe$_2$ is electron-like. The absence of an electron-like Hall signature in the single-phase AlB$_2$-type window provides no evidence for a dominant parallel-conduction channel from LaGe$_2$.

Figure~\ref{fig:3}f shows the longitudinal resistance $R_\mathrm{xx}$ versus temperature for films of fixed Ag/La ratio ($\approx$~0.20) with thickness ranging from 6.4 to 29.0~nm. These films remain well below the quantum resistance $h/e^2$, although the 6.4~nm film shows an upturn in resistance as the temperature is lowered, consistent with localization. Plotting 1/$R_\mathrm{xx}$ against thickness at 2~K and 300~K (Fig.~\ref{fig:3}g) permits an extrapolation with a positive thickness intercept, from which we extract a critical thickness of $\approx$5.3~nm below which conduction is suppressed. This critical thickness is larger than the $\approx$2~nm structurally disordered region directly resolved at the film/substrate interface by cross-sectional STEM (Fig.~\ref{fig:2}a), which we suspect contributes negligibly to transport, although the present measurements do not establish this. An equally plausible picture is that conduction is progressively suppressed by disorder-driven localization throughout the film as thickness decreases, without requiring a spatially distinct inactive region. The $\theta$--$2\theta$ peak positions shift only modestly across the thickness series, indicating a small thickness dependence of the average out-of-plane lattice parameter, and no secondary-phase peaks are detected (Fig.~S4).

\subsection{Superconductivity in AlB$_2$-type films}
\label{sec:supercon}

\begin{figure}
  \centering
  \includegraphics[width=1\linewidth]{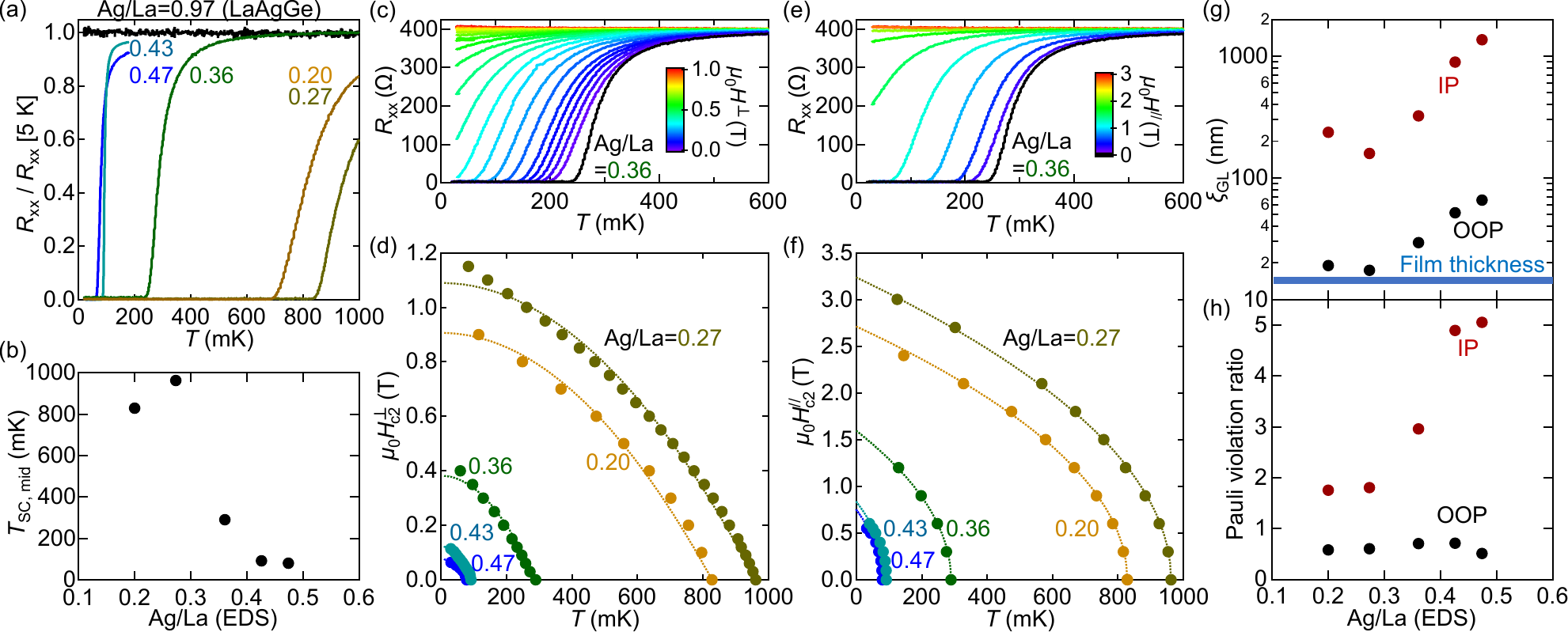}
  \caption{\textbf{Composition (Ag/La) dependence of the superconducting transition and its enhancement relative to the Pauli-field estimate.}
  (a) Normalized resistance $R/R(5\,\mathrm{K})$ near the superconducting transition for the single-phase AlB$_2$-type composition series (Ag/La = 0.20--0.43) and the adjacent Ag-rich phase-boundary sample (Ag/La = 0.47), referenced against the parent LaAgGe phase (Ag/La = 0.97).
  (b) Midpoint superconducting transition temperature $T_{\mathrm{SC,mid}}$ versus Ag/La (EDS).
  Temperature dependence of (c) $R_\mathrm{xx}$ for a representative film (Ag/La = 0.36) under out-of-plane field ($\mu_0H^{\perp}$, 0--1~T),
  (d) the out-of-plane upper critical field $\mu_0H_{\mathrm{c}2}^{\perp}(T)$, extracted from resistance-criterion curves as in (c), for films with Ag/La = 0.20, 0.27, 0.36, 0.43, and 0.47, together with Ginzburg--Landau (GL) fits (Supporting Information, Section~S2.2),
  (e) $R_\mathrm{xx}$ for the same representative film (Ag/La = 0.36) under in-plane field ($\mu_0H^{\parallel}$, 0--3~T), and
  (f) the in-plane upper critical field $\mu_0H_{\mathrm{c}2}^{\parallel}(T)$, extracted from resistance-criterion curves as in (e), together with thin-film GL fits.
  (g) Ag/La (EDS) dependence of the Ginzburg--Landau coherence length $\xi_{GL}$ for the in-plane and out-of-plane configurations. The horizontal line indicates the film thickness.
  (h) Ag/La (EDS) dependence of the ratio $\mu_0H_{\mathrm{c}2}/\mu_0H_\mathrm{P}$ (with $\mu_0H_\mathrm{P} = 1.86\,T_{\mathrm{SC,mid}}$) for the in-plane and out-of-plane configurations.
  }
  \label{fig:4}
\end{figure}
\begin{figure}
  \centering
  \includegraphics[width=0.95\linewidth]{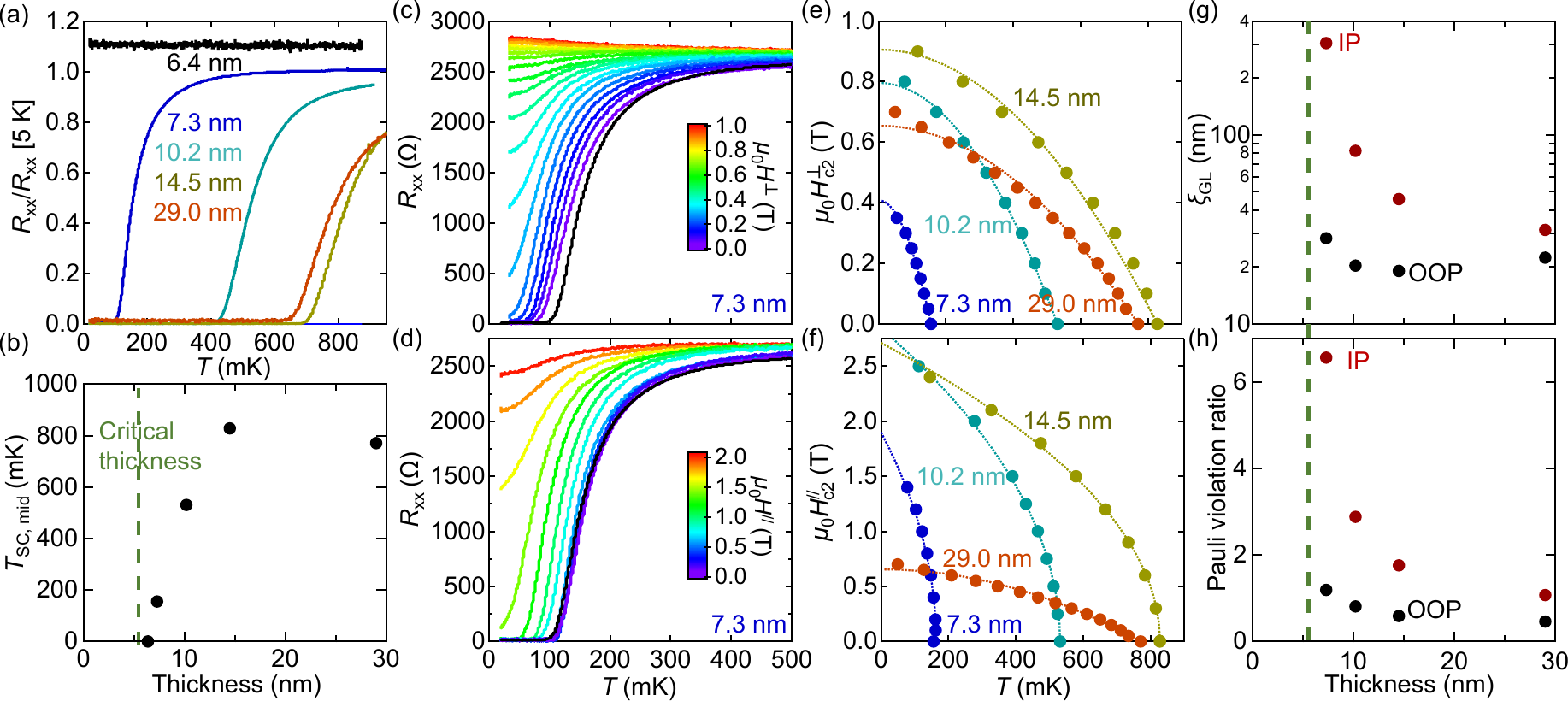}
  \caption{\textbf{Film-thickness dependence of the superconducting transition, upper critical field, and its enhancement relative to the Pauli-field estimate.}
  (a) Normalized resistance $R/R(5\,\mathrm{K})$ near the superconducting transition for the film-thickness series at fixed, Ag-deficient composition (Ag/La~$\approx$~0.20).
  (b) Midpoint superconducting transition temperature $T_{\mathrm{SC,mid}}$ versus film thickness. The dotted line marks the $\approx$5.3~nm critical thickness extracted in Figure~\ref{fig:3}g.
  Temperature dependence of (c) $R_\mathrm{xx}$  under out-of-plane field ($\mu_0H^{\perp}$, 0--1~T), and
  (d) $R_\mathrm{xx}$ under in-plane field ($\mu_0H^{\parallel}$, 0--2~T) for a representative film (7.3~nm).
  (e) out-of-plane upper critical field $\mu_0H_{\mathrm{c}2}^{\perp}(T)$, extracted from resistance-criterion curves as in (c), for films of thickness 7.3, 10.2, 14.5, and 29.0~nm, together with GL fits, and
  (f) in-plane upper critical field $\mu_0H_{\mathrm{c}2}^{\parallel}(T)$, extracted from resistance-criterion curves as in (d), together with thin-film GL fits.
  Thickness dependence of (g) the Ginzburg--Landau coherence length $\xi_\mathrm{GL}$ for the in-plane and out-of-plane configurations, and
  (h) $\mu_0H_{\mathrm{c}2}/\mu_0H_\mathrm{P}$ ratio for the in-plane and out-of-plane configurations.
  }
  \label{fig:5}
\end{figure}

The structural conversion to the AlB$_2$-type structure creates a distinct low-temperature electronic state: superconductivity emerges across a range of Ag/La ratios (Fig.~\ref{fig:4}) and film thicknesses (Fig.~\ref{fig:5}), whereas stoichiometric polar $P6_3mc$ LaAgGe remains metallic. The composition-series films in Fig.~\ref{fig:4} are 14--17~nm thick, well above the thickness at which localization suppresses conduction. Their superconducting critical temperature $T_{\mathrm{SC,mid}}$, defined at 50\% of the normal-state resistance, peaks near 1~K at Ag/La~$\approx$~0.27 (Fig.~\ref{fig:4}a,b). The upper critical field $H_{\mathrm{c}2}$ is strongly anisotropic and composition dependent for out-of-plane (Fig.~\ref{fig:4}c,d) and in-plane (Fig.~\ref{fig:4}e,f) field orientations. Fits to Ginzburg--Landau (GL) expressions appropriate to the two field orientations parameterize this response, and the extracted coherence lengths exceed the film thicknesses for every sample (Fig.~\ref{fig:4}g), placing the superconductivity in the two-dimensional limit. Relative to the weak-coupling Pauli estimate $\mu_0H_\mathrm{P}=1.86\,T_{\mathrm{SC,mid}}$, the in-plane $H_{\mathrm{c}2}$ is enhanced by a factor of 1.8 (Ag/La~=~0.20) to 5.0 (Ag/La~=~0.43) within the single-phase window and by a comparable factor in the Ag/La~=~0.47 phase-boundary sample (Fig.~\ref{fig:4}h). 
Possible contributions to the enhanced $H_{\mathrm{c}2}^{\parallel}$, including reduced orbital pair breaking and spin--orbit effects, are considered in the Discussion and Supporting Information.

The superconducting characteristics are also thickness-dependent, as illustrated in Fig.~\ref{fig:5}. Upon decreasing film thickness at a fixed Ag/La ratio of approximately 0.2, all films superconduct except the thinnest, at 6.4~nm. Tracking $T_{\mathrm{SC,mid}}$ as a function of thickness in Fig.~\ref{fig:5}b shows a rapid decline as the critical thickness is approached. The temperature-dependent superconducting transitions under out-of-plane and in-plane applied fields are shown in Fig.~\ref{fig:5}c,d for the thinnest superconducting film (7.3~nm), with their field dependencies summarized for all films in Fig.~\ref{fig:5}e,f. The complete field evolution of the 7.3~nm transition and the dependence of the extracted $H_{\mathrm{c}2}^{\parallel}(T)$ on the 10\%, 50\%, and 90\% resistance criteria are shown in Fig.~S9. The coherence length obtained through GL analysis is plotted in Fig.~\ref{fig:5}g, with the Pauli-limit ratio plotted in Fig.~\ref{fig:5}h. For the superconducting films thinner than 29.0~nm (7.3--14.5~nm), the out-of-plane-field-derived coherence length exceeds the film thickness, consistent with two-dimensional superconductivity in these thinner films; the 29.0~nm film does not satisfy this criterion. The Pauli-limit ratio increases monotonically as the film thickness decreases, consistent with progressive suppression of orbital depairing for an in-plane field.\cite{Dorrian2026_PairBreaking}

\section{Discussion}

The synthesis we present illustrates a general advantage of sequential transformations in multicomponent films; precursor epitaxy and target-phase composition can be established in separate steps, an approach that has also proved effective for other metastable epitaxial materials.\cite{GutierrezLlorente2024_Topochemical, Zhou2024_Epitaxial} In the case explored here, direct growth does not access the AlB$_2$-type phase, plausibly because it requires not only Ag removal but also sufficient lattice mobility to reconstruct the ordered, polar Ag/Ge sublattice into a planar honeycomb arrangement. Decreasing $P_\mathrm{Ag}/P_\mathrm{Ge}$ during growth expands the LaAgGe out-of-plane repeat as the films become more Ag deficient and eventually produces LaAgGe+LaGe$_x$ rather than the AlB$_2$-type phase (Fig.~S1). These as-grown films nevertheless retain longer out-of-plane repeats than the LaAgGe points in the annealed series without transforming. Post-growth annealing is therefore the essential kinetic step that enables the AlB$_2$-type reconstruction after the epitaxial template has formed. At the Ag-poor boundary, further annealing produces AlB$_2$-type+LaGe$_2$ rather than extending the solid solution continuously (Fig.~S2). The AlB$_2$-type phase therefore has a bounded homogeneity range, consistent with the narrow window ($x\approx0.25$--0.375) reported for bulk La(Au$_x$Ge$_{1-x}$)$_2$.\cite{Peterson2023_Twists} Separating epitaxial template formation from final composition selection suggests a transferable strategy for volatile, multicomponent intermetallics whose target phases compete within narrow stability windows. A tabulated comparison with reported bulk AlB$_2$-type $RE_2TX_3$ superconductors is provided in Table~S4.

The atomic-resolution results indicate local inversion-symmetry-breaking distortions within a structure that is centrosymmetric on average by diffraction. Related separations between average and local inversion symmetry have been reported in centrosymmetric superconductors, with consequences ranging from local structural distortions to hidden spin polarization.\cite{Griffitt2023_Local, Wu2017_Hidden} The measured Ag--Ge displacements indicate spatial variation in the magnitude and direction of the local structural distortion. Local FFT analysis reveals regions that retain the parent-like $c$-axis unit-cell doubling and regions in which the doubling is absent (Fig.~S6), providing additional evidence for spatially heterogeneous local structure. Together, these observations are consistent with local inversion-symmetry-breaking distortions within the diffraction-averaged $P6/mmm$ structure.

The increase of $H_{\mathrm{c}2}^{\parallel}/H_\mathrm{P}$ with decreasing film thickness is consistent with reduced orbital pair breaking for an in-plane field.\cite{Dorrian2026_PairBreaking} Because the composition-series films have similar nominal thicknesses of 14--17~nm, however, their strong composition dependence cannot be attributed to nominal film thickness alone. Spin--orbit coupling may also enhance $H_{\mathrm{c}2}^{\parallel}$ by suppressing the paramagnetic depairing channel, either through spin--orbit scattering or through local Rashba-type spin--orbit coupling associated with local inversion-symmetry breaking.\cite{Werthamer1966_Temperature, Frigeri2004_Superconductivity, Maruyama2012_Locally} The present measurements do not establish spin--orbit coupling or the local structural distortions as the microscopic origin of the enhancement. Other possible contributions, including effective superconducting thickness, disorder, electronic diffusivity, deviations from the weak-coupling Pauli-field estimate, and multiband effects, are discussed in the Supporting Information, Section~S2.3.

\section{Conclusions}
\label{sec:conclusion}

We demonstrate that post-growth annealing in the La--Ag--Ge ternary system stabilizes metastable AlB$_2$-type superconducting films within a bounded homogeneity range. Atomic-resolution imaging reveals local deviations from the planar AlB$_2$-type configuration that are consistent with inversion-symmetry-breaking structural distortions. Composition and thickness tune the lattice parameters, normal-state transport, and superconducting response of the converted films.
The broader advance is a synthesis--structure--property framework in which solid-state dealloying accesses a metastable quantum film, atomic-scale imaging reveals local structural distortions consistent with inversion-symmetry breaking that are hidden from the average structure, and composition and nanoscale thickness provide independent control parameters. This framework opens a route to investigating how local symmetry textures and dimensional confinement shape electronic states in volatile multicomponent intermetallic films.

\section{Methods}
\label{sec:methods}

\subsection{Film Growth and Annealing}
Before growth, Al$_2$O$_3$(0001) substrates were annealed at 1500~$^\circ$C for 10~min by CO$_2$ laser heating to promote stepped, atomically flat surfaces.\cite{Kim2025_HighTemperature, Glotzer2026_Thermallya} La--Ag--Ge films were grown by molecular-beam epitaxy (MBE). The chamber base pressure was approximately $10^{-10}$~mbar. Elemental effusion cells operated at approximately 1550, 840, and 1130~$^\circ$C for La, Ag, and Ge, respectively, and the Al$_2$O$_3$ substrate temperature was $T_\mathrm{g}=600~^\circ$C. Growth times of 35--180~min yielded a growth rate of approximately 0.16~nm/min. Beam-equivalent pressures (BEPs) were measured with an ionization gauge; elemental fluxes were calibrated using an \textit{in situ} quartz crystal monitor and cross-checked by X-ray reflectivity (XRR) on elemental witness films deposited at room temperature.\cite{Ohno2026_Anisotropic} Typical precursor growth used Ag/Ge and Ge/La flux ratios of approximately 1.1 and 0.45, respectively. The as-grown films were annealed \textit{in situ} under an Ag overpressure at $T_\mathrm{a}=800$--900~$^\circ$C for a total time $t_\mathrm{a}=10$--40~min, either as a single hold or as a sequence of temperature steps. The final EDS-measured Ag/La ratio and the corresponding phase and lattice parameters observed by XRD depend jointly on the precursor BEP ratio $P_\mathrm{Ag}/P_\mathrm{La}$, the annealing temperature $T_\mathrm{a}$, and the annealing time $t_\mathrm{a}$. In selected cases, the endpoint was identified from stabilization of the reflection high-energy electron diffraction (RHEED) intensity and streak shape. After annealing, the films were capped \textit{in situ} at room temperature with approximately 5~nm of amorphous Ge.

\subsection{Structural and Compositional Characterization}
X-ray diffraction (XRD) was performed using a Rigaku SmartLab diffractometer with Cu K$\alpha$ radiation. $\theta$--$2\theta$ scans, rocking curves, reciprocal-space maps (RSMs) around asymmetric reflections, and azimuthal $\varphi$-scans were used to determine phase content, lattice parameters, and epitaxial relationships. Film thickness was determined from XRR fringe analysis. Composition was quantified using a ZEISS 1550VP field-emission scanning electron microscope equipped with an Oxford Instruments X-Max silicon drift detector for energy-dispersive X-ray spectroscopy (EDS). Spectra were analyzed using Oxford Instruments AZtec software with standardless background modeling. All reported Ag/La composition ratios are EDS-measured values. As-grown, unannealed LaAgGe films gave Ag/La~$\approx$~0.97 as an internal check against the nominal 1:1:1 composition and the XRD phase assignment. Samples were mounted on carbon tape for EDS, producing the C signal in Fig.~S5; the Al and O signals arise from the Al$_2$O$_3$ substrate. Consequently, these spectra do not independently resolve interface-specific oxygen incorporation.
Crystal-structure renderings in Fig.~\ref{fig:1}a and the structural models used in Fig.~\ref{fig:2}b,c were generated from reported crystallographic information files for polar LaAgGe ($P6_3mc$) and AlB$_2$-type La(Au$_{0.278}$Ge$_{0.723}$)$_2$ ($P6/mmm$). \cite{Pecharskii1991_ChemInform, Peterson2023_Twists} The reported lattice parameters and fractional atomic coordinates were retained. For the AlB$_2$-type La--Ag--Ge representations, Au was replaced by Ag on the mixed honeycomb site. The resulting models serve as structural references and do not represent crystallographic refinements of the present films.

\subsection{Electron Microscopy}
Cross-sectional lamellae were prepared by focused-ion-beam (FIB) lift-out using an FEI Helios NanoLab 600 DualBeam. A protective cap was deposited before milling; the lamellae were thinned with 5~kV Ga ions and polished at 2~kV. High-angle annular dark-field scanning transmission electron microscopy (HAADF-STEM) was performed using a Thermo Scientific Themis Z S/TEM operated at 300~kV with a 30~mrad probe semi-convergence angle and a 64--200~mrad detector range. Each image was formed by registering and averaging a series of 20 fast-scan frames (2048~$\times$~2048 pixels; 200~ns pixel dwell time). Z-sensitive HAADF contrast distinguishes relatively Ag-rich and Ge-rich honeycomb-site columns. Atomic-column positions were determined from one-dimensional HAADF-STEM intensity profiles extracted along local $c$ and averaged over 1.36~\AA-wide strips parallel to local $a$. Adjacent Ag-rich and Ge-rich profiles were jointly fitted with Gaussian peaks sharing a common width and separate linear backgrounds using custom Python code based on NumPy and SciPy. The signed $c$-axis Ag--Ge displacement was determined from the relative fitted positions of adjacent Ag-rich and Ge-rich columns using the reference configurations defined in Fig.~\ref{fig:2}c.

\subsection{Electrical Transport and Upper-Critical-Field Analysis}
Electrical transport above approximately 2~K was measured in a Quantum Design DynaCool Physical Property Measurement System (PPMS) equipped with a 9~T superconducting magnet. Measurements below 1~K were performed in a Leiden Cryogenics dilution refrigerator (DR) with a base temperature of approximately 20~mK and a two-axis 9--3~T vector magnet. Rectangular film pieces were contacted with Al wires in a four-terminal geometry and measured using a lock-in amplifier with a 100--500~nA excitation at 10--20~Hz. The in-plane magnetic field was aligned using the established vector-magnet rotation procedure.\cite{Llanos2026_Fieldinduced, Dorrian2026_PairBreaking} Longitudinal and Hall signals were obtained by symmetrizing and antisymmetrizing the field-dependent data, respectively. The measured longitudinal resistance $R_\mathrm{xx}^{\mathrm{meas}}$ was converted to the sheet resistance $R_\mathrm{xx}$ by applying the inverse aspect ratio. The Hall coefficient $R_\mathrm{H}$ was obtained from the linear-in-field slope of $R_\mathrm{yx}(B)$. The effective sheet carrier density was calculated using the single-band expression $n_\mathrm{2D}=1/(eR_\mathrm{H})$, where $e$ is the elementary charge. The Hall mobility was obtained from the sheet conductance $G_\square=1/R_\mathrm{xx}$ as $\mu=G_\square/(n_\mathrm{2D}e)$, and the conductivity was estimated as $\sigma_\mathrm{xx}=G_\square/t$ using the nominal film thickness $t$. The residual resistance ratio was defined as $\mathrm{RRR}=R_\mathrm{xx}(300~\mathrm{K})/R_\mathrm{xx}(2~\mathrm{K})$. Only $n_\mathrm{2D}$ is reported as the carrier-density parameter. For the single-band mean-free-path estimate, however, a spherical three-dimensional Fermi surface was assumed and $n_\mathrm{2D}/t$ entered the expression only as an intermediate quantity, giving $\ell=(\hbar\mu/e)\left(3\pi^2n_\mathrm{2D}/t\right)^{1/3}$ where $\hbar$ is the reduced Planck constant, and $n_\mathrm{2D}$, $t$, and $\mu$ are expressed in SI units. Because the electronic structure may be multiband and composition-dependent, $n_\mathrm{2D}$ and $\ell$ are treated as effective single-band transport parameters.

Upper critical fields $H_{\mathrm{c}2}(T)$ were extracted from fixed-field $R_\mathrm{xx}(T)$ sweeps using the 50\% resistance criterion, where the normal-state resistance $R_0$ is the maximum resistance before transition onset. The 10\% and 90\% criteria were additionally used to represent the transition width, and the resulting criterion dependence of $T_{\mathrm{SC,mid}}$ and $H_{\mathrm{c}2}^{\parallel}(T)$ is summarized in Fig.~S9. The Ginzburg--Landau fitting expressions and their assumptions are provided in the Supporting Information, and the fit parameters are summarized in Table~S3. For each film in the composition series, XRD, EDS, and transport measurements were performed on the same sample piece; the quoted Ag/La ratio is therefore a single-sample EDS determination, with an estimated uncertainty of approximately $\pm$0.02--0.03 from counting statistics and the standardless quantification procedure.


\section{Acknowledgments}
We acknowledge funding provided by the Institute for Quantum Information and Matter, an NSF Physics Frontiers Center (NSF Grant PHY-2317110), and the Gordon and Betty Moore Foundation's EPiQS Initiative (Grant number GBMF10638).
We acknowledge the support of JSPS Overseas Research Fellowships.
This material is based upon work supported by the National Science Foundation Graduate Research Fellowship Program under Grant No. 2139433 (RD, VS).
Any opinions, findings, and conclusions or recommendations expressed in this material are those of the author(s) and do not necessarily reflect the views of the National Science Foundation.
Electron microscopy was performed at the Center for Electron Microscopy and Analysis (CEMAS) at The Ohio State University.
\section*{References}
\bibliography{bib}
\clearpage

\begin{center}
\Large\bfseries Supplementary Materials
\end{center}

\setcounter{section}{0}
\setcounter{subsection}{0}
\setcounter{figure}{0}
\setcounter{table}{0}
\setcounter{equation}{0}
\renewcommand{\thesection}{S\arabic{section}}
\renewcommand{\thefigure}{S\arabic{figure}}
\renewcommand{\thetable}{S\arabic{table}}
\renewcommand{\theequation}{S\arabic{equation}}
\renewcommand{\theHsection}{S\arabic{section}}
\renewcommand{\theHsubsection}{S\arabic{section}.\arabic{subsection}}
\renewcommand{\theHfigure}{S\arabic{figure}}
\renewcommand{\theHtable}{S\arabic{table}}
\renewcommand{\theHequation}{S\arabic{equation}}

\section{Structure}
\label{sec:S1}
\subsection{Growth-series phase selection}
\label{sec:S1_growth}

\begin{figure}[h]
\centering
\includegraphics[width=0.6\linewidth]{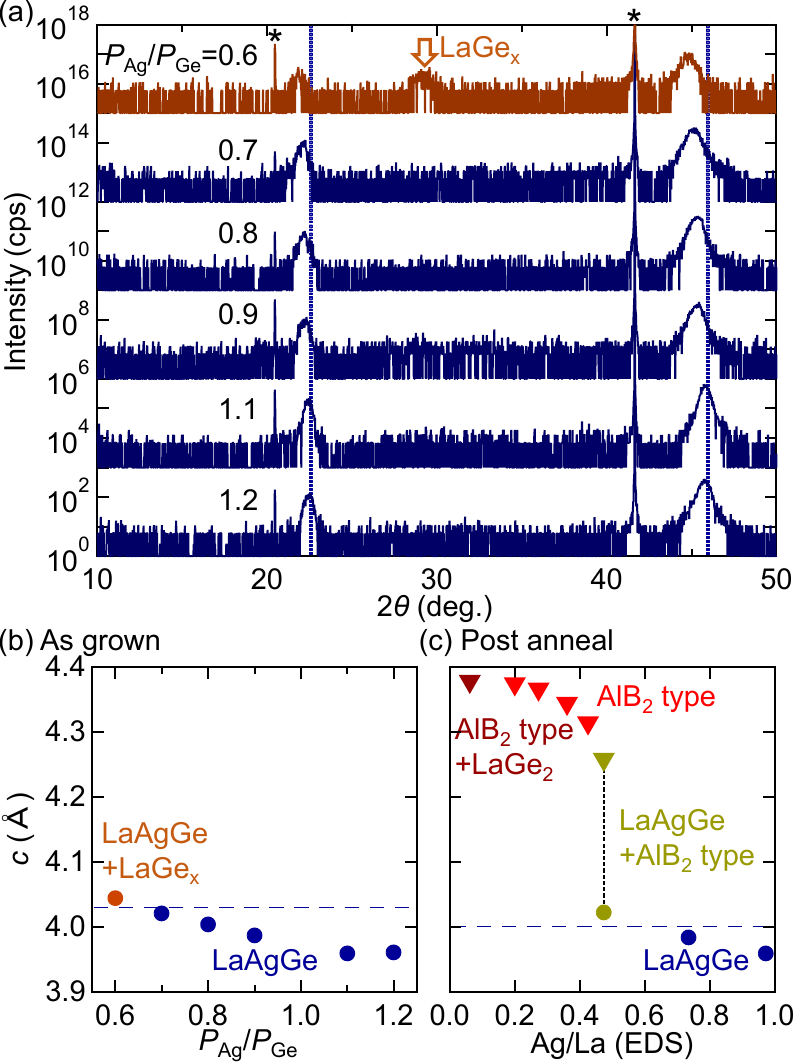}
\caption{\textbf{Ag deficiency alone does not stabilize the AlB$_2$-type phase during growth.} (a) Out-of-plane $\theta$--$2\theta$ scans for as-grown films deposited with $P_\mathrm{Ag}/P_\mathrm{Ge}=0.6$--1.2. A LaGe$_x$ reflection appears at $P_\mathrm{Ag}/P_\mathrm{Ge}=0.6$, but no AlB$_2$-type reflection is observed. Al$_2$O$_3$ substrate peaks are marked with an asterisk. (b) LaAgGe out-of-plane repeat versus $P_\mathrm{Ag}/P_\mathrm{Ge}$, showing expansion with increasing Ag deficiency. (c) Annealing-driven out-of-plane repeat versus EDS-measured Ag/La, reproducing main-text Figure~1e.}
\label{sfig:asgrown_diagram}
\end{figure}

Film growth, annealing, capping, thickness calibration, and composition quantification are described in the main-text Methods. Figure~\ref{sfig:asgrown_diagram} provides the additional growth-flux series used to distinguish Ag deficiency during deposition from the post-growth annealing transformation.

Figure~\ref{sfig:asgrown_diagram} separates the effect of growth-flux-controlled Ag deficiency from that of post-growth annealing. In the as-grown series, decreasing $P_\mathrm{Ag}/P_\mathrm{Ge}$ systematically increases the LaAgGe out-of-plane repeat as the films become more Ag deficient. Films deposited with $P_\mathrm{Ag}/P_\mathrm{Ge}=0.7$--1.2 remain LaAgGe, while the most Ge-rich condition ($P_\mathrm{Ag}/P_\mathrm{Ge}=0.6$) additionally contains LaGe$_x$; no AlB$_2$-type phase is observed anywhere in the as-grown series. Panel~(c), which presents the same annealing-driven evolution as main-text Figure~1e, shows the XRD-derived out-of-plane repeat as a function of post-anneal Ag/La composition. Importantly, the as-grown LaAgGe films in panel~(b) retain longer out-of-plane repeats than the LaAgGe points in panel~(c), yet do not transform into the AlB$_2$-type phase. Ag deficiency alone is therefore insufficient to stabilize the AlB$_2$-type structure during deposition; post-growth annealing is essential for the transformation.

\subsection{X-ray diffraction (XRD)}
\label{sec:S1_xrd}
The XRD instrumentation and measurement geometries are described in the main-text Methods. This section presents the extended composition- and thickness-dependent diffraction datasets.

\begin{figure}[h]
\centering
\includegraphics[width=0.6\linewidth]{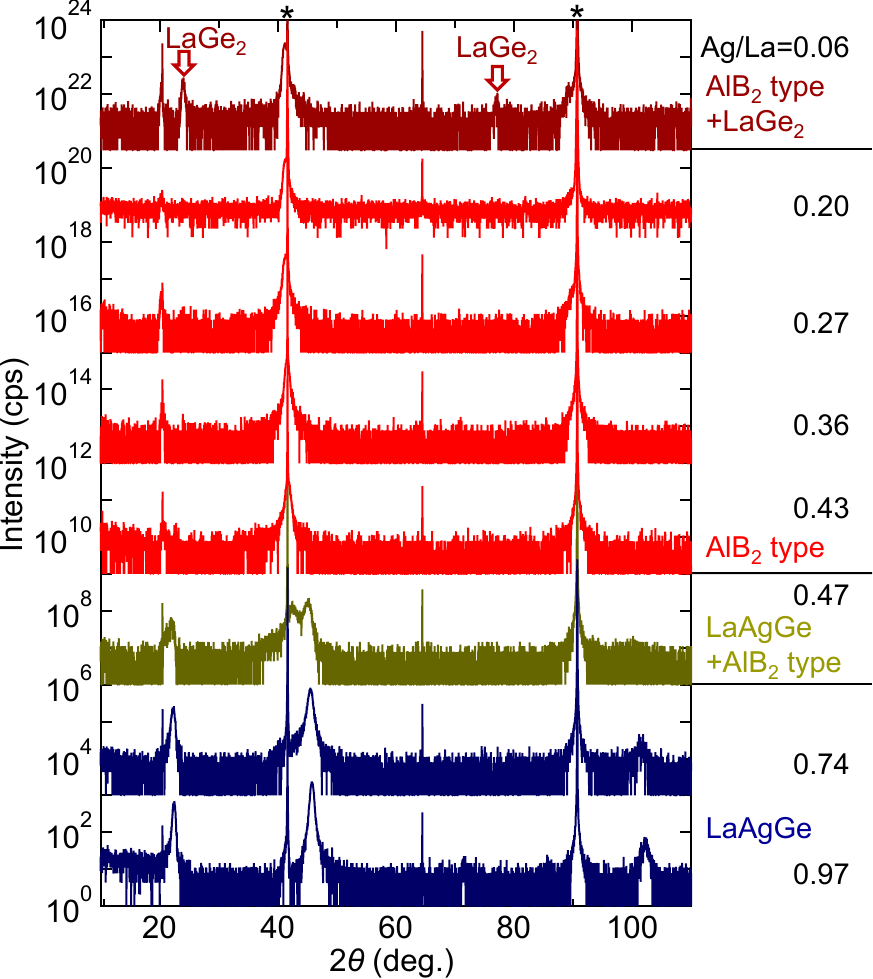}
\caption{\textbf{Wide-angle XRD across the composition series.} Out-of-plane $\theta$--$2\theta$ scans over $2\theta=10$--110$^\circ$ for Ag/La~=~0.06, 0.20, 0.27, 0.36, 0.43, 0.47, 0.74, and 0.97. The patterns show the evolution from the AlB$_2$-type+LaGe$_2$ two-phase region at Ag/La~=~0.06, through the single-phase AlB$_2$-type region, to LaAgGe+AlB$_2$-type at Ag/La~=~0.47 and LaAgGe at Ag/La~=~0.74 and 0.97. LaGe$_2$ reflections are marked for Ag/La~=~0.06. Al$_2$O$_3$ substrate peaks are marked with an asterisk.}
\label{sfig:xrd_composition}
\end{figure}
Wide-angle $\theta$--$2\theta$ scans over $2\theta=10$--110$^\circ$ confirm the phase assignments across the composition series (Figure~\ref{sfig:xrd_composition}): the complete family of (00$l$) AlB$_2$-type or LaAgGe reflections is present, with no detectable secondary-phase peaks outside the two-phase boundary compositions discussed in the main text.

\begin{figure}[h]
\centering
\includegraphics[width=1\linewidth]{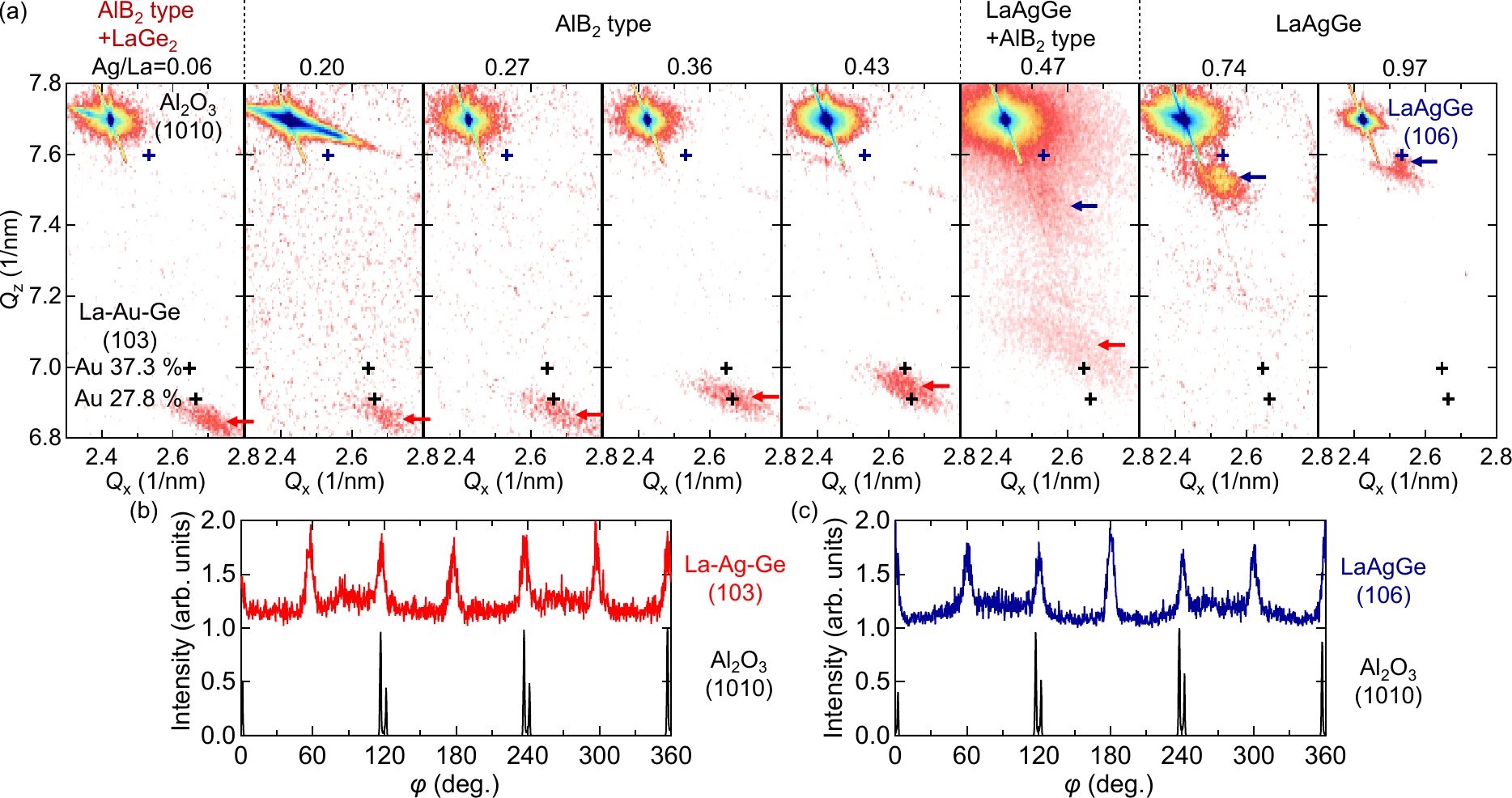}
\caption{\textbf{Reciprocal space maps and azimuthal scans across the composition series.} (a) Reciprocal space maps around the Al$_2$O$_3$ $(1010)$ substrate reflection and the LaAgGe (106) or AlB$_2$-type (103) film reflection for Ag/La~=~0.06--0.97. The phase assignments are indicated above the maps; arrows mark the film reflections. (b) Azimuthal $\varphi$-scan of the AlB$_2$-type La--Ag--Ge (103) reflection together with the Al$_2$O$_3$ $(1010)$ substrate reflection. (c) Corresponding $\varphi$-scan of the LaAgGe (106) and Al$_2$O$_3$ $(1010)$ reflections. The sixfold film peaks establish a single, well-defined in-plane epitaxial registry.}
\label{sfig:rsm_all}
\end{figure}

\begin{figure}[h]
\centering
\includegraphics[width=1\linewidth]{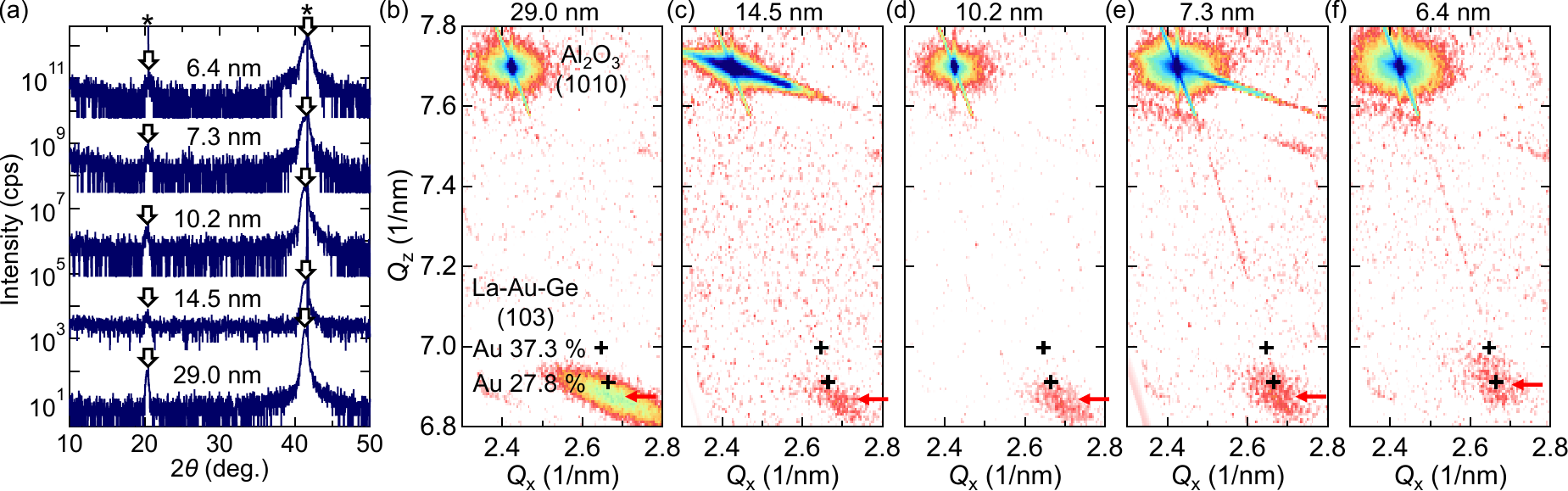}
\caption{\textbf{XRD across the thickness series.} (a) Out-of-plane $\theta$--$2\theta$ scans over $2\theta=10$--50$^\circ$ for films 6.4, 7.3, 10.2, 14.5, and 29.0~nm thick at fixed Ag-deficient composition. Open arrows mark the AlB$_2$-type film reflections. Al$_2$O$_3$ substrate peaks are marked with an asterisk. (b--f) Reciprocal space maps around the Al$_2$O$_3$ $(1010)$ substrate reflection and the AlB$_2$-type (103) film reflection for thicknesses of (b) 29.0, (c) 14.5, (d) 10.2, (e) 7.3, and (f) 6.4~nm.}
\label{sfig:xrd_wide_thickness}
\end{figure}

Reciprocal space maps (RSMs) around the Al$_2$O$_3$ $(1010)$ substrate reflection and the LaAgGe (106)/AlB$_2$-type (103) film reflection are shown for the full composition series in Figure~\ref{sfig:rsm_all}a and for the full thickness series in Figure~\ref{sfig:xrd_wide_thickness}b--f. Across the composition series, the reciprocal-space maps show composition-dependent changes in film-reflection shape and width. Because the present dataset does not include a systematic quantitative line-shape analysis, these changes are not used to extract a correlation length or structural-heterogeneity fraction. Across the thickness series, the film reflection retains a well-defined spot at fixed in-plane momentum transfer, indicating unchanged epitaxial registry over this thickness range.

Azimuthal $\varphi$-scans of the AlB$_2$-type (103)/Al$_2$O$_3$ $(1010)$ reflection pair and the LaAgGe (106)/Al$_2$O$_3$ $(1010)$ reflection pair (Figure~\ref{sfig:rsm_all}b,c, respectively) each display six equally spaced film peaks at a fixed azimuthal offset from the substrate peaks, confirming sixfold in-plane symmetry and a single, well-defined epitaxial orientation relationship with the Al$_2$O$_3$(0001) substrate throughout the composition series.

For the thickness series, $\theta$--$2\theta$ scans over the $2\theta=10$--50$^\circ$ range (Figure~\ref{sfig:xrd_wide_thickness}a) show only modest shifts in the (00$l$) AlB$_2$-type peak position across the full thickness range studied, with no detectable secondary-phase peaks. These data indicate a small thickness dependence of the average out-of-plane lattice parameter and show that the AlB$_2$-type phase is retained without detectable phase separation over the measured thickness range.

\clearpage
\subsection{Electron microscopy and composition analysis}
\label{sec:S1_stem}
\begin{figure}[h]
\centering
\includegraphics[width=1\linewidth]{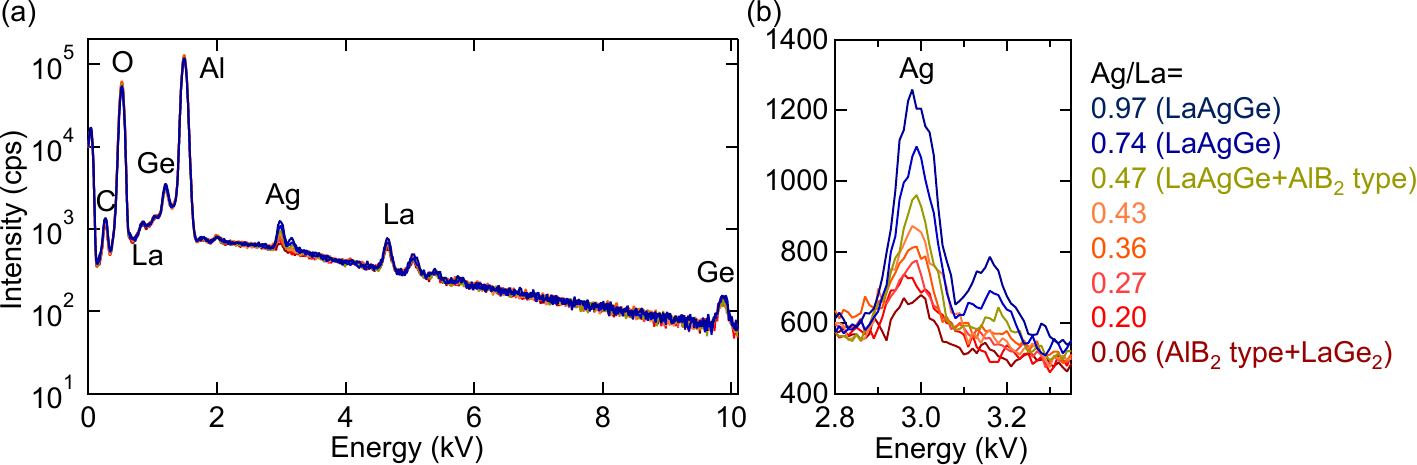}
\caption{\textbf{EDS spectra across the composition series.} (a) Wide-range spectra showing the labeled C, O, Al, Ge, Ag, and La peaks for films with EDS-measured Ag/La~=~0.06--0.97. (b) Enlarged view of the Ag $L$-line region. Colors and labels identify the XRD phase assignments.}
\label{sfig:eds}
\end{figure}

\begin{figure}[h]
\centering
\includegraphics[width=1\linewidth]{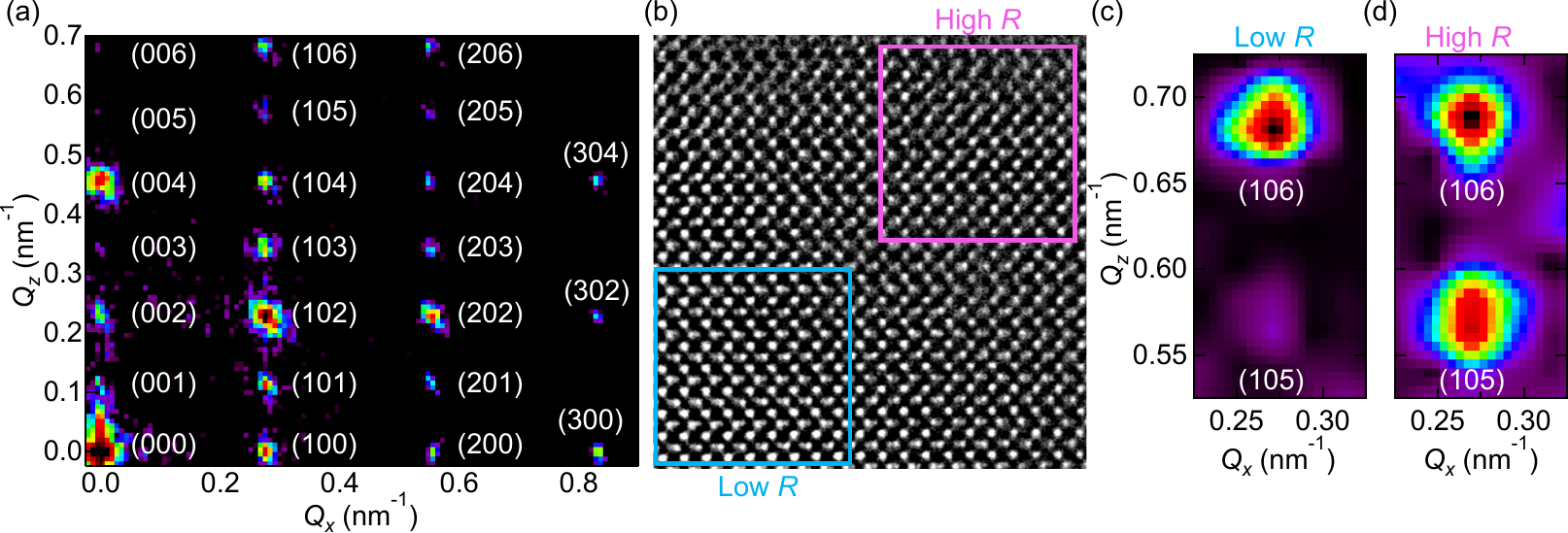}
\caption{\textbf{Full-field and local FFT analysis of the $c$-axis periodicity.}
(a) Indexed fast Fourier transform (FFT) calculated from the full AlB$_2$-type region visible in the HAADF-STEM image shown in main-text Figure~2a. All FFT reflections are indexed using the LaAgGe parent cell, whose $c$ axis is doubled relative to the AlB$_2$-type cell. The full-field FFT power is displayed on a logarithmic color scale.
(b) Higher-magnification HAADF-STEM image shown in main-text Figure~2b, with regions labeled ``Low-$R$'' (blue) and ``High-$R$'' (pink), where $R$ denotes the ratio of the (105) FFT intensity to the (106) FFT intensity.
(c,d) Local FFTs calculated from the (c) Low-$R$ and (d) High-$R$ regions. In contrast to the full-field of view FFT in (a), the local FFT intensities are normalized to their respective (106) reflections and displayed over the same reciprocal-space range using a common linear color scale. The (105) reflection is strongly suppressed in the Low-$R$ region but remains visible in the High-$R$ region, consistent with spatial variation in the doubled-$c$-sensitive FFT contrast.}
\label{sfig:stem_fft}
\end{figure}

Electron-microscopy specimen preparation, imaging conditions, and energy-dispersive X-ray spectroscopy (EDS) quantification are described in the main-text Methods. Figure~\ref{sfig:eds} documents the composition-series spectra, and Figure~\ref{sfig:stem_fft} provides the supplementary fast Fourier transform (FFT) analysis of the scanning transmission electron microscopy (STEM) images.

The wide-range spectra in Figure~\ref{sfig:eds}a confirm the presence of C, O, Al, Ge, Ag, and La. The C signal originates from the carbon tape used to mount the samples, whereas the Al and O signals originate predominantly from the Al$_2$O$_3$ substrate. Across the film series, the C, O, Al, Ge, and La peak intensities remain nearly unchanged within experimental variation, while the Ag peak changes systematically. The enlarged Ag $L$-line region in Figure~\ref{sfig:eds}b shows that the Ag intensity is highest in the as-grown film with Ag/La~$\approx$~0.97 and decreases after annealing as the measured Ag/La ratio is reduced. The spectra therefore support net Ag loss as the primary composition change detected by EDS during the annealing-driven transformation.

Figure~\ref{sfig:stem_fft} shows spatial variation of the $c$-axis periodicity within the imaged region of the converted film. The full-field FFT in panel~(a) retains finite intensity at the (105) superstructure position even though the macroscopic XRD map shows no measurable intensity there after conversion. This difference is consistent with the much smaller sampling area and greater local sensitivity of STEM rather than a contradiction between the two measurements.

Panels~(b--d) compare representative Low-$R$ and High-$R$ regions. The (105) reflection is strongly suppressed in the Low-$R$ region but remains visible in the High-$R$ region, demonstrating spatial variation in the doubled-$c$ superstructure intensity. The labels describe relative FFT contrast and do not define discrete structural phases or a phase fraction.

Within the analyzed atomic-resolution field of view, the measured signed Ag--Ge displacements span both directions relative to the planar AlB$_2$-type reference (main-text Figure~2d). These relative displacements, together with the spatial variation in the (105) FFT contrast, are consistent with local inversion-symmetry-breaking structural distortions within the diffraction-averaged centrosymmetric phase. The analysis characterizes the range of local configurations observed within this field of view and is not used to define discrete structural populations or their fractions.

\clearpage
\section{Transport and Superconductivity}
\label{sec:S2}

\subsection{Normal-state transport data}
\label{sec:S2_transport}

Transport instrumentation, contact geometry, excitation conditions, field alignment, data symmetrization, and Hall-parameter definitions are described in the main-text Methods. This section provides the complete field-dependent datasets and derived parameter tables.

\begin{figure}[th]
\centering
\includegraphics[width=0.98\linewidth]{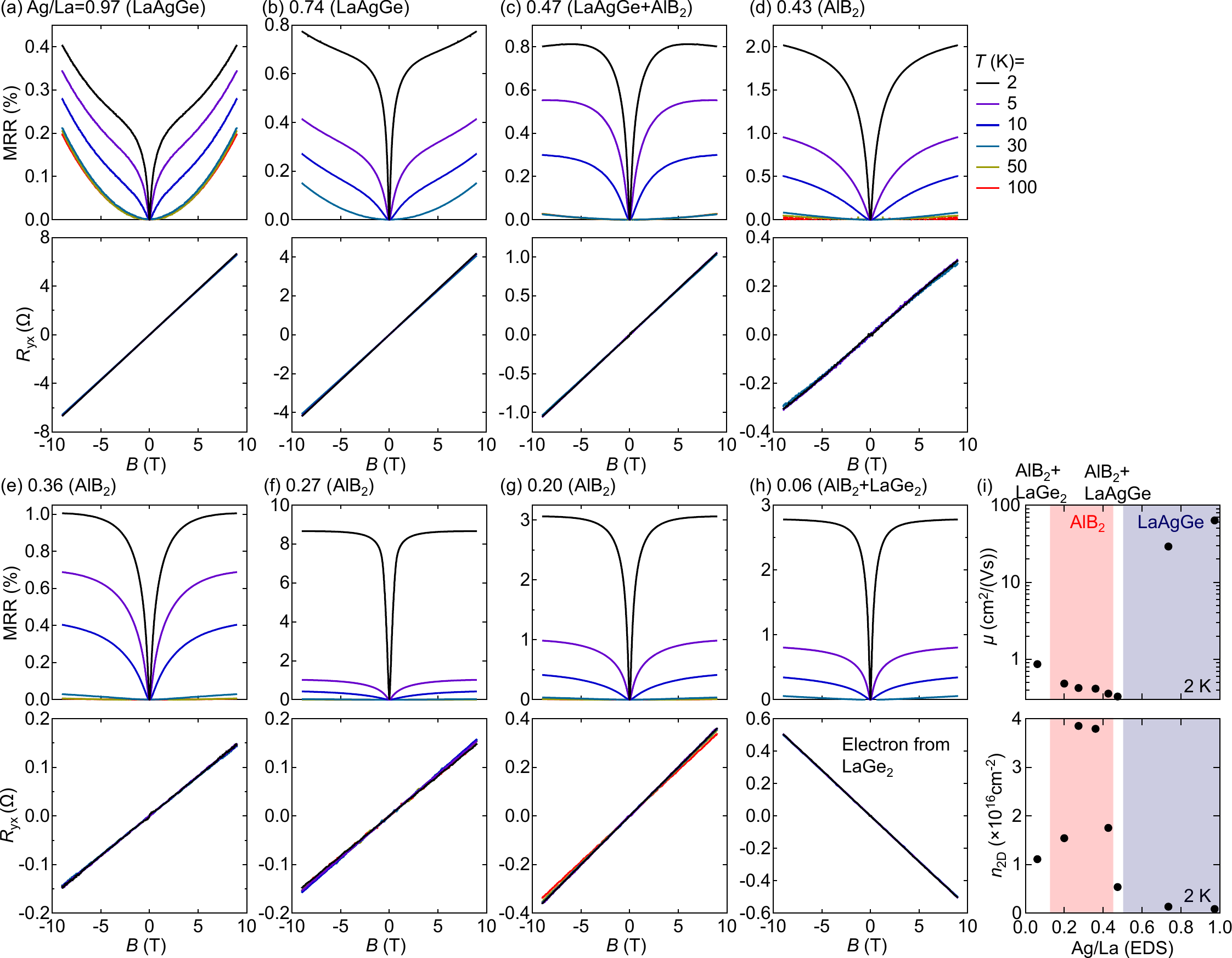}
\caption{\textbf{Magnetoresistance and Hall response across the LaAgGe--AlB$_2$ composition series.} (a--h) Magnetoresistance ratio $\mathrm{MRR}=[R_\mathrm{xx}(B)-R_\mathrm{xx}(0)]/R_\mathrm{xx}(0)$, where $R_\mathrm{xx}$ is the longitudinal resistance (upper subpanel), and Hall resistance $R_\mathrm{yx}(B)$ (lower subpanel) measured at 2, 5, 10, 30, 50, and 100~K for Ag/La~=~0.97, 0.74, 0.47, 0.43, 0.36, 0.27, 0.20, and 0.06, respectively. LaAgGe and the single-phase AlB$_2$-type films exhibit a hole-like Hall response, whereas the LaGe$_2$-containing Ag/La~=~0.06 film in (h) exhibits an electron-like response. (i) Effective single-band Hall mobility $\mu$ (upper subpanel) and sheet carrier density $n_\mathrm{2D}$ (lower subpanel) at 2~K versus EDS-measured Ag/La. Vertical dotted lines mark the approximate compositional regions.}
\label{sfig:mr_composition}
\end{figure}

\begin{table}[ht]
\centering
\small
\caption{Normal-state transport parameters measured at 2~K across the Ag/La composition series. The residual resistance ratio is defined as $\mathrm{RRR}=R_\mathrm{xx}(300~\mathrm{K})/R_\mathrm{xx}(2~\mathrm{K})$. The sheet carrier densities $n_\mathrm{2D}$ are effective single-band values obtained directly from the Hall slope. For the mean-free-path estimate $\ell$, $n_\mathrm{2D}/t$ enters only as an intermediate quantity under the spherical three-dimensional Fermi-surface assumption described in the main-text Methods.}
\label{tab:transport_composition}
\begin{tabular}{ccccrc}
\hline
Ag/La
& RRR
& $n_\mathrm{2D}$ ($10^{16}$~cm$^{-2}$)
& $\mu$ (cm$^2$/(V\,s))
& $\sigma_\mathrm{xx}$ ($\Omega^{-1}$\,cm$^{-1}$)
& $\ell$ (nm) \\
\hline
0.97 & 1.10 & 0.084 & 63.3 & 3702 & 9.2  \\
0.74 & 1.04 & 0.13  & 29.2 & 3676 & 5.5 \\
0.47 & 1.03 & 0.54  & 0.33  & 170  & 0.10 \\
0.43 & 1.03 & 1.75   & 0.36  & 625  & 0.16 \\
0.36 & 1.07 & 3.79   & 0.42  & 1690 & 0.25 \\
0.27 & 1.08 & 3.85   & 0.43  & 1783 & 0.26 \\
0.20 & 1.04 & 1.54   & 0.49  & 823  & 0.22 \\
0.06 & 1.08 & 1.11   & 0.87  & 1083 & 0.35 \\
\hline
\end{tabular}
\end{table}

Figure~\ref{sfig:mr_composition} summarizes the field-dependent transport across the LaAgGe--AlB$_2$ composition series. The magnetoresistance is positive and generally decreases with increasing temperature. In several superconducting films, the enhanced low-field magnetoresistance at 2~K can include field-induced suppression of superconducting fluctuations or precursor superconducting transport. It therefore should not be interpreted solely as intrinsic normal-state magnetoresistance or as evidence for weak antilocalization. At higher temperatures, where superconducting contributions are suppressed, the normal-state magnetoresistance remains comparatively modest.

The LaGe$_2$-containing Ag/La~=~0.06 film shows an electron-like Hall response, opposite in sign to the hole-like response throughout the single-phase AlB$_2$-type composition range. The absence of an electron-like Hall signature in the single-phase films provides no evidence for a dominant LaGe$_2$ parallel-conduction channel, although a smaller contribution cannot be excluded. This result is consistent with the absence of detectable LaGe$_2$ peaks in the XRD patterns of the single-phase AlB$_2$-type films.

The 6.4~nm film, which shows no measurable superconducting transition, yields an effective single-band mean-free-path estimate of $\ell\approx0.04$~nm, below a typical interatomic spacing. This unphysical scale shows that the spherical single-band semiclassical model is no longer quantitatively valid in this regime; it is used only as a qualitative indicator of very strong disorder.

\begin{table}[ht]
\centering
\caption{Normal-state transport parameters measured at 2~K for the thickness series at a fixed Ag-deficient composition (Ag/La~$\approx$~0.20). The sheet carrier densities $n_\mathrm{2D}$ are effective single-band values obtained directly from the Hall slope. For the mean-free-path estimate $\ell$, $n_\mathrm{2D}/t$ enters only as an intermediate quantity under the spherical three-dimensional Fermi-surface assumption described in the main-text Methods.}
\label{tab:transport}
\begin{tabular}{rcccccc}
\hline
$t$ (nm) & RRR & $n_\mathrm{2D}$ ($10^{16}$~cm$^{-2}$) & $\mu$ (cm$^2$/(V\,s)) & $\sigma_\mathrm{xx}$ ($\Omega^{-1}$\,cm$^{-1}$) & $\ell$ (nm) & Superconducting \\
\hline
6.4  & 0.81 & 1.8 & 0.06 & 257  & 0.04 & No  \\
7.3  & 0.95 & 1.5 & 0.31  & 1025 & 0.17 & Yes \\
10.2 & 0.99 & 0.86 & 0.44  & 592  & 0.18 & Yes \\
14.5 & 1.04 & 1.5 & 0.49  & 823  & 0.22 & Yes \\
29.0 & 1.07 & 4.0 & 0.79  & 1741 & 0.38 & Yes \\
\hline
\end{tabular}
\end{table}

The effective sheet carrier density $n_\mathrm{2D}$ varies nonmonotonically with thickness, decreasing from $4.0\times10^{16}$~cm$^{-2}$ at 29.0~nm to $0.86\times10^{16}$~cm$^{-2}$ at 10.2~nm before increasing in the two thinner films. Because $n_\mathrm{2D}$ is a single-band transport parameter, these data do not identify the microscopic origin of the trend; sample-to-sample composition, disorder, and multiband transport can all contribute.

\begin{figure}[th]
\centering
\includegraphics[width=0.75\linewidth]{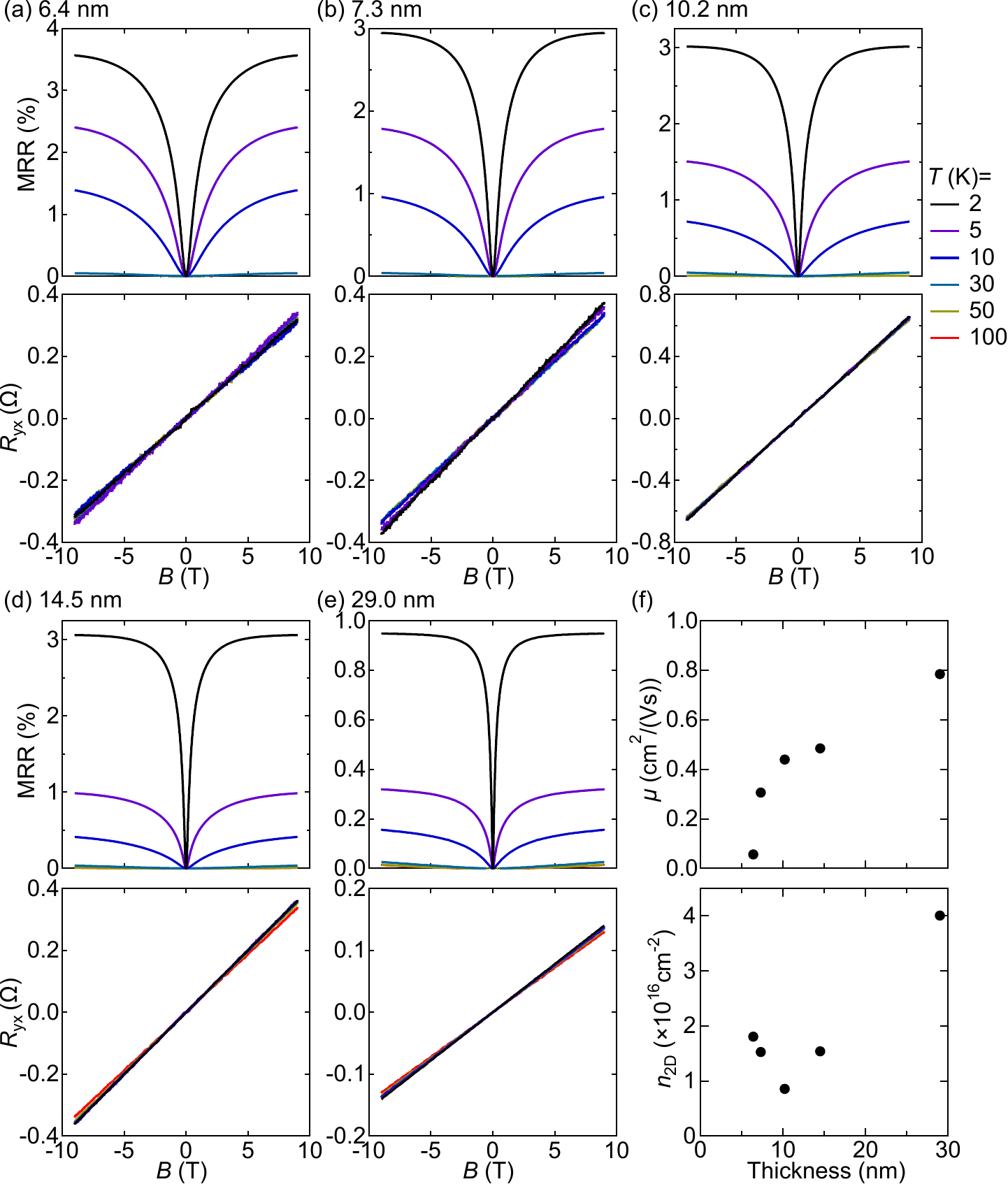}
\caption{\textbf{Magnetoresistance and Hall response across the film-thickness series.} (a--e) Magnetoresistance ratio $\mathrm{MRR}$ (upper subpanel) and Hall resistance $R_\mathrm{yx}(B)$ (lower subpanel) measured at 2, 5, 10, 30, 50, and 100~K for films with thicknesses of 6.4, 7.3, 10.2, 14.5, and 29.0~nm, respectively, at fixed Ag-deficient composition (Ag/La~$\approx$~0.20). All films show a hole-like Hall response. (f) Effective single-band Hall mobility $\mu$ (upper subpanel) and sheet carrier density $n_\mathrm{2D}$ (lower subpanel) at 2~K versus nominal film thickness.}
\label{sfig:mr_thickness}
\end{figure}

Figure~\ref{sfig:mr_thickness} shows the corresponding field-dependent transport across the thickness series. The magnetoresistance remains positive and generally decreases with increasing temperature, while the Hall response remains hole-like at every thickness. Panel~(f) shows that the effective mobility generally increases with thickness, whereas the effective sheet carrier density varies nonmonotonically.

\clearpage
\subsection{Ginzburg--Landau analysis of the upper critical field}
\label{sec:S2_hc2}
We parameterize the temperature dependence of the resistively defined upper-critical-field scale using Ginzburg--Landau (GL)-motivated expressions.\cite{Tinkham1996_Introduction} These fits provide phenomenological extrapolations of the zero-temperature field scales and the in-plane GL coherence length; they are not thermodynamic determinations of $H_{\mathrm{c}2}$ and are not used to identify a microscopic pair-breaking mechanism.

\textbf{Out-of-plane field:}
\begin{equation}
\mu_0H_{\mathrm{c}2}^{\perp}(T)
= \frac{\Phi_0}{2\pi\xi_{ab}^2(0)}
\left[1-\left(\frac{T}{T_\mathrm{SC}}\right)^2\right].
\label{eq:hc2_perp_gl}
\end{equation}

\textbf{In-plane field, Tinkham thin-film limit with fixed thickness $d$:}
\begin{equation}
\mu_0H_{\mathrm{c}2}^{\parallel}(T)
= \frac{\sqrt{12}\,\Phi_0}{2\pi\xi_{ab}(0)d}
\sqrt{1-\frac{T}{T_\mathrm{SC}}}.
\label{eq:hc2_parallel_tinkham}
\end{equation}

\begin{table}[bh]
\centering
\caption{GL parameters extracted from $H_{\mathrm{c}2}(T)$. Perpendicular-field data were fitted using the standard GL expression, and in-plane data using the fixed-thickness thin-film GL expression. The listed $\xi_\mathrm{GL}^{\perp}(0)$ values are obtained from the perpendicular-field fits. Composition-series values are approximate, whereas thickness-series values are obtained from the fits. Ag/La~=~0.47 is a phase-boundary, two-phase sample (main text) and is included for reference alongside the single-phase composition series (Ag/La~=~0.20--0.43).}
\label{tab:glfits}
\begin{tabular}{lcccc}
\hline
Sample & $T_{\mathrm{SC,mid}}$ (K) & $\mu_0H_{\mathrm{c}2}^{\perp}(0)$ (T) & $\xi_\mathrm{GL}^{\perp}(0)$ (nm) & $\mu_0H_{\mathrm{c}2}^{\parallel}(0)$ (T) \\
\hline
\multicolumn{5}{l}{Composition series ($\sim 15$~nm)} \\
Ag/La = 0.47 & 0.08 & 0.08 & 65.6 & 0.75 \\
Ag/La = 0.43 & 0.09 & 0.12 & 51.7 & 0.84 \\
Ag/La = 0.36 & 0.29 & 0.38 & 29.4 & 1.60 \\
Ag/La = 0.27 & 0.96 & 1.09 & 17.4 & 3.24 \\
Ag/La = 0.20 & 0.83 & 0.91 & 19.1 & 2.71 \\
\hline
\multicolumn{5}{l}{Thickness series (Ag/La~$\sim$~0.20)} \\
7.3~nm  & 0.16 & 0.34 & 31.0 & 1.90 \\
10.2~nm & 0.53 & 0.80 & 20.3 & 2.84 \\
14.5~nm & 0.83 & 0.91 & 19.1 & 2.71 \\
29.0~nm & 0.77 & 0.66 & 20.4 & 1.53 \\
\hline
\end{tabular}
\end{table}

Equation~\ref{eq:hc2_parallel_tinkham} is the two-dimensional thin-film result derived for a field parallel to the film.\cite{Tinkham1963_Fluxoid} Here $\Phi_0$ is the magnetic flux quantum, $\xi_{ab}(0)$ is the zero-temperature in-plane GL coherence length inferred from the perpendicular-field fit, $d$ is the measured film thickness, and $T_\mathrm{SC}$ is the fitted zero-field transition temperature. Because the strict thin-film condition is not uniformly satisfied by all samples, the in-plane expression is used only as a phenomenological parameterization. These fits do not uniquely separate orbital and paramagnetic pair breaking; consequently, no spin--orbit-scattering time or other microscopic scattering parameter is inferred from them. The extracted parameters are listed in Table~\ref{tab:glfits}.

\begin{figure}[h]
\centering
\includegraphics[width=0.8\linewidth]{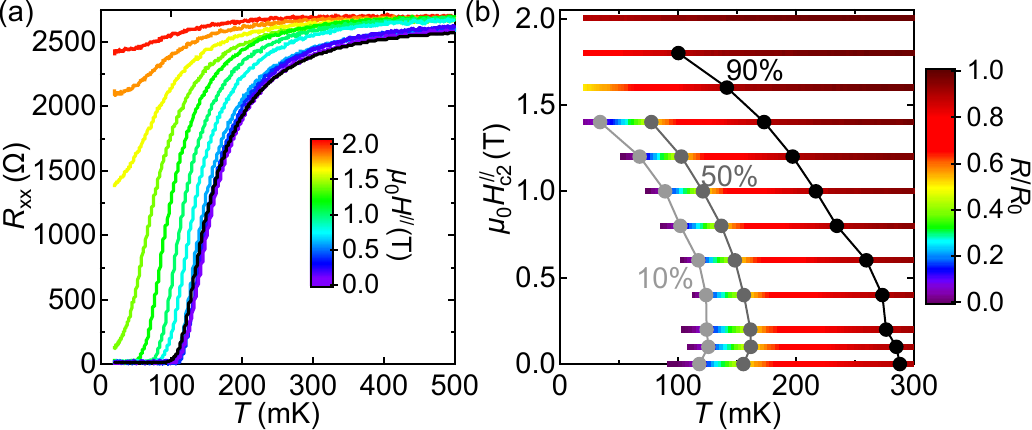}
\caption{\textbf{Field evolution of the broad superconducting transition in the 7.3~nm film.} (a) Longitudinal resistance $R_\mathrm{xx}(T)$ at fixed in-plane magnetic fields $\mu_0H^{\parallel}=0$--2.0~T. (b) The same data represented by the normalized resistance $R/R_0$ as a function of temperature and in-plane field. Black and gray symbols connected by lines trace the 90\%, 50\%, and 10\% resistance criteria; the 50\% criterion is used for the main-text $H_{\mathrm{c}2}^{\parallel}(T)$ analysis.}
\label{sfig:field_enhanced}
\end{figure}

The reported field scales use the 50\% resistance criterion defined in the main-text Methods. Figure~\ref{sfig:field_enhanced} compares the 10\%, 50\%, and 90\% contours of $R/R_0$ for the broad transition in the 7.3~nm film, explicitly showing the criterion dependence of the inferred $H_{\mathrm{c}2}^{\parallel}(T)$. For the midpoint values in Table~\ref{tab:glfits}, $\mu_0H_{\mathrm{c}2}^{\parallel}(0)=1.90$~T and $T_{\mathrm{SC,mid}}=0.16$~K give $H_{\mathrm{c}2}^{\parallel}/H_\mathrm{P}^{\mathrm{BCS}}\approx6.4$. This ratio should be interpreted together with the transition width and criterion dependence rather than as a criterion-independent thermodynamic quantity.

Across the single-phase window, $T_{\mathrm{SC,mid}}$ rises from 0.83~K at Ag/La~=~0.20 to 0.96~K at Ag/La~$\approx$~0.27 and then falls to 0.09~K at Ag/La~=~0.43. The concurrent reduction of $T_{\mathrm{SC,mid}}$, $\sigma_\mathrm{xx}$, and RRR near the phase boundaries establishes a correlation but does not distinguish composition, disorder, and structural heterogeneity as its cause.

\subsection{Extended discussion}
\label{sec:S2_discussion}
Table~\ref{tab:literature} compares the present films with reported AlB$_2$-type or AlB$_2$-derived superconductors. In the cited reports, superconductivity was measured in polycrystalline bulk material prepared primarily by arc melting or direct solid-state synthesis, with post-annealing where required. Although In-flux-grown Yb$_2$PdGe$_3$ single crystals were reported, they were used for structural characterization; superconductivity was measured on a polycrystalline sample.

\begin{table}[h]
\centering
\caption{Comparison with reported AlB$_2$-type or AlB$_2$-derived superconductors. The superconducting properties in the cited prior reports were measured on polycrystalline bulk samples. Separately prepared In-flux-grown Yb$_2$PdGe$_3$ single crystals were used only for structural characterization. The present epitaxial La--Ag--Ge films enable orientation-resolved upper-critical-field measurements. A dash indicates that a value was not reported.}
\label{tab:literature}

\scriptsize
\setlength{\tabcolsep}{5pt}
\renewcommand{\arraystretch}{0.95}

\resizebox{\textwidth}{!}{%
\begin{tabular}{lllllr}
\hline
Material
& Sample form
& $T_\mathrm{SC}$ (K)
& $\mu_0H_{\mathrm{c}2}$ (T)
& Synthesis
& Ref. \\
\hline

La$_2$NiGe$_3$
& Polycrystalline bulk
& 0.394
& ---
& Arc melting + annealing
& \citenum{Chen2012_Superconductivity} \\

Y$_2$NiGe$_3$
& Polycrystalline bulk
& 0.58
& ---
& Arc melting + annealing
& \citenum{Chen2012_Superconductivity} \\

Y$_2$PtGe$_3$
& Polycrystalline bulk
& 3.3
& ---
& Arc melting
& \citenum{Kito2002_Superconductivity} \\

Y$_2$PdGe$_3$
& Polycrystalline bulk
& 3.0
& ---
& Arc melting + annealing
& \citenum{Majumdar2001_Observation} \\

Y$_2$Pd$_{1.25}$Ge$_{2.75}$
& Polycrystalline bulk
& 2.72
& 2.94
& Arc melting
& \citenum{Swiatek2024_Detailed} \\

Y$_2$PdGe$_{2.7}$Si$_{0.3}$
& Polycrystalline bulk
& 3.55
& ---
& Arc melting + annealing
& \citenum{Ghosh2003_Superconductivity} \\

Y$_2$PdGe$_{1.6}$Si$_{1.4}$
& Polycrystalline bulk
& 2.05
& ---
& Arc melting + annealing
& \citenum{Ghosh2003_Superconductivity} \\

Y$_2$Pd$_{1-x}$Pt$_x$Ge$_3$
& Polycrystalline bulk
& 3.2 -- 2.15
& ---
& Arc melting + annealing
& \citenum{Iyer2007_Superconducting} \\

Yb$_2$PdGe$_3$
& Polycrystalline bulk
& 4.0
& ---
& Direct solid-state synthesis
& \citenum{Freccero2023_Flux} \\

CaCu$_{0.33}$Si$_{1.67}$
& Polycrystalline bulk
& 2.3
& ---
& Arc melting + annealing
& \citenum{Hor2006_Superconductivity} \\

CaCu$_{0.25}$Si$_{1.75}$
& Polycrystalline bulk
& 3.1
& ---
& Arc melting + annealing
& \citenum{Hor2006_Superconductivity} \\

CaCu$_{0.25}$Si$_{1.70}$
& Polycrystalline bulk
& 3.2
& ---
& Arc melting + annealing
& \citenum{Hor2006_Superconductivity} \\

\hline

\shortstack[l]{La--Ag--Ge\\\strut}
& \shortstack[l]{Epitaxial film\\\strut}
& \shortstack[l]{0.08--0.96\\\strut}
&
\shortstack[l]{0.08--1.09 ($\perp$)\\0.75--3.24 ($\parallel$)}
&
\shortstack[l]{MBE +\textit{in situ}\\solid-state dealloying}
& 
\shortstack[l]{This\\work}
\\

\hline
\end{tabular}%
}
\end{table}

Unlike the polycrystalline bulk samples used for the superconducting measurements summarized in Table~\ref{tab:literature}, the epitaxial film geometry of the present samples permits separate measurements for fields perpendicular and parallel to the film. The resulting in-plane values reach $\mu_0H_{\mathrm{c}2}^{\parallel}=0.75$--3.24~T and exceed the weak-coupling BCS Pauli-field estimate for several samples. The Clogston--Chandrasekhar estimate, $\mu_0H_\mathrm{P}^{\mathrm{BCS}}=1.86\,T_\mathrm{SC}$ in tesla for $T_\mathrm{SC}$ in kelvin, assumes weak-coupling spin-singlet pairing and $g=2$.\cite{Clogston1962_Upper,Chandrasekhar1962_Maximum} This orientation-resolved comparison motivates the possible pair-breaking mechanisms considered below.

The GL analysis parameterizes $H_{\mathrm{c}2}(T)$ but does not identify the microscopic origin of the apparent exceedance of the Pauli estimate. Within the single-phase composition series, $H_{\mathrm{c}2}^{\parallel}/H_\mathrm{P}^{\mathrm{BCS}}$ increases from approximately 1.8 at Ag/La~=~0.20 to 5.0 at Ag/La~=~0.43 even though $\mu_0H_{\mathrm{c}2}^{\parallel}(0)$ decreases from approximately 2.71 to 0.84~T. The ratio grows because $T_{\mathrm{SC,mid}}$, and therefore $H_\mathrm{P}^{\mathrm{BCS}}$, decreases more rapidly than the absolute in-plane field scale toward the Ag-rich boundary.

In the thickness series, the increasing ratio with decreasing thickness is consistent with reduced orbital pair breaking for an in-plane field.\cite{Dorrian2026_PairBreaking} The composition-series films have similar nominal thicknesses, but inhomogeneity near the Ag-rich phase boundary could reduce the effective superconducting thickness and produce a similar orbital effect.\cite{Klemm1975_Theory, Dorrian2026_PairBreaking} The present measurements do not resolve the spatial distribution of the superconducting condensate, so this possibility cannot be quantified.
Other contributions include disorder-dependent diffusivity, spin--orbit impurity scattering, multiband superconductivity, and local antisymmetric spin--orbit coupling permitted by the noncentrosymmetric distortions.\cite{Kapitulnik1985_Anderson, Werthamer1966_Temperature, Gurevich2003_Enhancement, Frigeri2004_Superconductivity, Maruyama2012_Locally, Griffitt2023_Local, Wu2017_Hidden} In addition, $H_\mathrm{P}=1.86\,T_\mathrm{SC}$ assumes weak-coupling singlet pairing and $g\approx2$, so changes in gap strength, electron--phonon renormalization, spin susceptibility, or effective $g$ factor would change the relevant paramagnetic scale.\cite{Clogston1962_Upper, Orlando1979_Critical} Because microscopy covers only one composition and no direct spin-sensitive measurement is available, the present data do not select among these mechanisms or establish a connection between local symmetry breaking and the enhanced field response.

\end{document}